\documentclass[twocolumn,aps,pra]{revtex4-1}
\usepackage{amsmath}    
\usepackage{graphicx}   
\usepackage{verbatim}   
\usepackage{color}      
\usepackage{mathrsfs}
\usepackage{amsfonts}
\usepackage{graphicx}   
\usepackage{verbatim}   
\usepackage{color}      
\usepackage{subfigure}  
\usepackage{hyphenat}
\usepackage[normalem]{ulem}
\usepackage{array}
\usepackage{epstopdf}

\begin{document}

\title{The importance of thermodynamics for molecular systems, and 
the importance of molecular systems for thermodynamics
}

\author{ Thomas E. Ouldridge      
}


\affiliation{
	Department of Bioengineering, Imperial College London, South Kensington Campus, London, SW7 2AZ, UK.
              \email{t.ouldridge@imperial.ac.uk}           
}


\begin{abstract}
Improved understanding of molecular systems has only emphasised the sophistication of  networks within the cell. Simultaneously, the advance of nucleic acid nanotechnology, a platform within which reactions can be exquisitely controlled, has made the development of artificial architectures and devices possible. Vital to this progress has been a solid foundation in the thermodynamics of molecular systems. In this pedagogical review and perspective, we discuss how  thermodynamics determines both the overall potential of molecular networks, and the minute details of design. We then argue that, in turn, the need to understand molecular systems is helping to drive the development of theories of thermodynamics at the microscopic scale.
\end{abstract}

\maketitle

\section{Introduction}
\label{sec:intro}
Thermodynamics was originally developed in the 19th Century, driven by the dawn of the industrial revolution \cite{Carnot1825}, and a desire to understand and optimise the extraction of useful work from engines. This work could be harnessed to pump water out of mines, or drive locomotives, for example. Although these machines were mechanical devices  powered by the flow of heat, the fundamental  source of the work was the chemical fuel - typically coal, initially. Chemical processes were incorporated into the framework of thermodynamics by Gibbs and others \cite{Gibbs1876}, leading to an understanding of the spontaneity of, and energy exchanged during, chemical reactions and phase changes. 

Thermodynamics as originally introduced was a theory based entirely on the interrelation of macroscopic observables, such as temperature, pressure, volume and energy. In the late 19th and early 20th centuries, the development of statistical mechanics provided a microscopic basis for the theory, explaining how these bulk properties emerge from microscopic system properties \cite{Boltzmann1896}. In the process, the concept of entropy -- the mysterious quantity whose increase is responsible for the thermodynamic arrow of time -- was made far more concrete, as a statistical measure of the uncertainty of the microscopic state of a system. The exploration of statistical mechanics  led to the development of theories of critical phenomena; explaining the exotic yet often universal behaviour of systems as they approach certain kinds of phase transitions \cite{Yeomans1992,Cardy1996}; statistical mechanics is also the fundamental tool underlying the field of molecular simulation \cite{Frenkel2001,Tuckerman2010}.

Statistical mechanics and thermodynamics are often introduced as the study of equilibrium states, in which there is no net tendency for the system to evolve over time unless driven from the outside. Nonetheless, from the earliest days, Boltzmann and others proposed theoretical descriptions for the evolution of non-equilibrium systems and their subsequent relaxation to equilibrium \cite{Boltzmann1896}. The developing field of stochastic thermodynamics, in which the probabilistic description underlying statistical mechanics is extended to describe trajectories of non-equilibrium systems through state space, has recently provided remarkable understanding of systems arbitrarily far from equilibrium \cite{Crooks1999,Seifert2005,Esposito2011,Jarzynski2011,Seifert2012}. 

Over the same period, our understanding of biomolecular systems has been transformed from complete ignorance to the ability to rationally design synthetic circuits and self-assembling architectures {\it in vitro} and {\it in vivo}. Indeed, although proteins were identified as far back as the early 19th Century \cite{Mulder1839,Teich1992}, their role as enzymes in living organisms was not demonstrated until 1926 \cite{Sumner1926}, and protein structures were first solved in 1958 \cite{Kendrew1958,Muirhead1963}. Similarly, the genetic information-carrying role of biological DNA was first demonstrated in 1944 \cite{Avery1944}, the double-helical structure was solved in 1953 \cite{Watson1953} and the central dogma of molecular biology {(that genes encoded in DNA are transcribed into RNA, and then translated into proteins)} was first stated in 1958 \cite{Crick1958,Crick1970}. Since then, through advances in crystallography, microscopy and other technologies, the molecular mechanisms of an enormous biochemical processes have been identified. Additionally, systems biology has shown how individual component reactions can combine to provide the complex behaviour exhibited by cells \cite{Alon2007} -- although we remain far from a full understanding of such sophisticated systems.

In the process of understanding some of the molecular complexity of the cell, we have shown that it contains microscopic analogues of the mechanical engines of the 19th Century. Molecular motors such as myosin consume chemical fuel to generate locomotive forces \cite{Howard2001}, and enzymatic pumps consume the same fuel to drive ions across membranes \cite{Nelson2004}. These membranes then act as capacitors that provide an alternative supply of power, like batteries for electric motors. 

A deep appreciation of natural biomolecular systems is worthwhile in and of itself, and it provides an important contribution to the advancement of medicine. But equally, this hard-won understanding has laid the groundwork for the engineering of artificial systems and devices. In synthetic biology, novel molecular circuitry is often built by connecting naturally-occuring or slightly mutated proteins via artificial transcriptional regulation pathways \cite{Baldwin2012}. At the same time, the molecular nanotechnology community has constructed systems based on artificial components, including non-biological DNA  and RNA sequences \cite{Chen2015} and even artificial proteins \cite{Hsia2016,Bale2016}. At the interface of these communities are those who combine the functionality of synthetic and natural components, both {\it in vivo} and {\it in vitro} \cite{Fujii2012,Green2014}. In aggregate, this work has produced remarkable results, including nanoscale self-assembly \cite{Rothemund06,Douglas09,Ke2012}, implementation of molecular computation and control architectures \cite{Winfree98,Seelig2006,Qian2011,Zechner2016} and repurposing of microbes for industry and healthcare \cite{Baldwin2012,Kitney2012,Chubukov2016}. 

In  this pedagogical perspective,  we will first discuss the basics of traditional chemical thermodynamics as it applies to biomolecular systems {(Sections~\ref{sec:fundamentals} and \ref{sec:free_energies}}).  We subsequently show how these ideas shape our understanding and design of functional molecular systems, both at a fundamental and a practical level, {in Sections~\ref{sec:self_assembly1} to \ref{sec:catalysis}}. {A particular focus will be common misconceptions or pitfalls that result from careless treatment of the underlying thermodynamics.}
Finally, we briefly discuss the emerging field of stochastic thermodynamics {in Section~\ref{sec:stoch_thermo}}. This extension of traditional thermodynamics to fluctucating, far-from equilibrium contexts finds its most natural application in the analysis of molecular systems. Indeed, we will then argue that  the very process of exploring abstract thermodynamic ideas in concrete biomolecular systems is in turn providing a deeper understanding of the fundamental thermodynamics.

\section{Fundamentals of classical statistical mechanics}
\label{sec:fundamentals}

\subsection{The partition function and thermodynamic quantities}
\label{sec:partition_fn}
{We will begin by considering the properties of an arbitrary, closed, equilibrium system, and then develop those ideas to arrive at the statistical mechanics of biochemical systems in particular. 
A closed system has fixed amounts of energy and matter. Nonetheless, a large closed system can access an enormous number of microstates specified by the positions and momenta $({\bf x}, {\bf p})$ of all the constituent degrees of freedom. In classical ({as opposed to quantum}) physics, these microstates are assumed to be present with a constant density $\rho$ throughout the available phase space defined by $({\bf x}, {\bf p})$ \cite{Frenkel2001}.}

{An experimenter that could perform the (impossible)  task of measuring the precise microstate $({\bf x}, {\bf p})$ of our system would not get a predictable value. This uncertainty can be quantified by $P({\bf x}, {\bf p})$, the probability per unit phase space volume of observing a microstate $({\bf x}, {\bf p})$ upon measurement. Initially, we are interested in characterising equilibrium systems, which exhibit no net flows between any pair of microstates and thus have $P({\bf x}, {\bf p})$ constant over time. The fundamental assumption of classical statistical mechanics  is that in thermodynamic equilibrium, there is the maximal possible uncertainty in the microstate $({\bf x}, {\bf p})$ \cite{Jaynes1957}. In other words,  all accessible (equal-energy)  {microstates} are equally probable in equilibrium \cite{Huang1987}. In the  rest of this Section, we further expand on the the properties of the equilibrium distribution that follow from this principle of equal {\it a priori} probability.}


In the molecular context, we are usually interested in a relatively small system $\sigma$ thermally connected to a much larger environment $\Sigma$, rather than a system in total  isolation. This larger environment might be the lab as a whole, or perhaps a water bath for elevated temperatures. In any case, we are typically not concerned with the details of this environment $\Sigma$, other than in its role as a source and sink of energy in the form of heat. Generally, we assume that the coupling is weak so that it is reasonable to separately consider the energies of $\Sigma$ and $\sigma$ \cite{Frenkel2001,Huang1987}.

The combined system of $\sigma+\Sigma$ remains a closed system, which can in principle reach equilibrium. However,
due to energy exchange between $\sigma$ and $\Sigma$, the energy of each component fluctuates  and microstates of $\sigma$ with different energies $E_\sigma({\bf x}, {\bf p})$ can be accessed. From applying the principle of equal {\it a priori} probability to the combined system of $\sigma+\Sigma$, it is possible to show that energy is shared such that  microstates of $\sigma$ are occupied with a probability density \cite{Frenkel2001,Tuckerman2010,Huang1987}
\begin{equation}
P^{\rm eq}_\sigma({\bf x}, {\bf p}) =  \rho\frac{\exp(-E_\sigma({\bf x}, {\bf p})/k_{\rm B}T)}{Z_\sigma},
\label{eq:Bmann}
\end{equation}
where the environmental heat bath sets the temperature $T$, and the partition function $Z_\sigma$ normalizes the distribution:
\begin{equation}
Z_\sigma = {\rho} \int {\mathrm d}{\bf x}\,{\mathrm d}{\bf p}  \exp(-E_\sigma({\bf x}, {\bf p})/k_{\rm B}T).
\label{eq:pn_fn}
\end{equation}
Here, $\rho$ is the constant density of microstates that cancels out during calculations, but ensures the correct dimensionality.
Eq.~\ref{eq:pn_fn} is the famous Boltzmann distribution \cite{Frenkel2001,Tuckerman2010,Huang1987}; its form arises from sharing energy between $\sigma$ and $\Sigma$ in such a way as to maximise overall uncertainty in the microstate. 

Atomistic models of molecular systems, such as AMBER and CHARMM \cite{Orozco2003,Cino2012}, are essentially semi-empirical energy functions $E_\sigma({\bf x}, {\bf p})$. Coarse-grained models, such as Martini and oxDNA \cite{Marrink2007,Doye2013}, are attempts to capture the behaviour of a reduced set of key degrees of freedom with a similar energy model. In either case, the models are typically too complex to be analysed directly. Instead, simulation is used to sample microstates of $\sigma$, allowing the equilibrium properties of the system to be inferred.

 The key thermodynamic quantities follow from the partition function. Firstly, the internal energy of $\sigma$ is a straightforward average over microsate energies \cite{Frenkel2001,Tuckerman2010,Huang1987}
\begin{equation}
U^{\rm eq}_\sigma =   \int_{{\bf x},{\bf p}} {\mathrm d}{\bf x},{\mathrm d}{\bf p}\,  E_\sigma({\bf x},{\bf p}) P_\sigma^{\rm eq}({\bf x},{\bf p}) = k_{\rm B} T^2 \frac{\partial}{\partial T} \ln Z_\sigma.
\label{eq:U}
\end{equation}

{The thermodynamic entropy is less obvious -- it is interpreted as a measure of the statistical uncertainty of the microstate distribution $P_\sigma^{\rm eq}({\bf x},{\bf p}) $ \cite{Jaynes1957}. This interpretation is made plausible because the second law states that the entropy of a closed system cannot decrease with time as it converges towards equilibrium;  and the principle of equal {\it a priori} probability that implies that the equilibrium distribution maximises uncertainty for a closed system.  Specifically, for an arbitrary discrete distribution  $P(y)$ over the variable $y$, the uncertainty in $y$ is described by the statistical entropy  \cite{Jaynes1957,Shannon1949}}
\begin{equation}
\mathcal{H}[P]= - \sum_y P(y) \ln P(y).
\label{eq:discrete entropy}
\end{equation}
{
Note that $\mathcal{H}[P] \geq 0$ is minimised when $y$ takes a single value with probability 1, and maximised if $p(y)$ is uniform \cite{Shannon1949}.

Allowing for continuous variables and introducing the constant $k_{\rm B}$ to connect to physical quantities, the equilibrium thermodynamic entropy of $\sigma$ is given by a similar expression:}
\begin{align}
S^{\rm eq}_\sigma &= - k_{\rm B}\int_{{\bf x},{\bf p}} {\mathrm d}{\bf x},{\mathrm d}{\bf p}\, P^{\rm eq}_\sigma({\bf x},{\bf p}) \ln ( P^{\rm eq}_\sigma({\bf x},{\bf p})/\rho) \nonumber \\
 & =  k_{\rm B} T \frac{\partial}{\partial T} \ln Z_\sigma  + k_{\rm B} \ln Z_\sigma.
\label{eq:entropy}
\end{align}
 This definition of entropy in Eq.~\ref{eq:entropy} is essentially the fundamental link between equilibrium statistical mechanics and macroscopic equilibrium thermodynamics. Finally, the free energy of $\sigma$  follows as \cite{Frenkel2001,Tuckerman2010,Huang1987}
\begin{equation}
F^{\rm eq}_\sigma = U^{\rm eq}_\sigma-TS^{\rm eq}_\sigma = -k_{\rm B}T \ln Z_\sigma.
\label{eq:F}
\end{equation}

The thermodynamic quantities listed above are the natural quantities of interest when we consider a system being manipulated from the outside, and thereby transitioning between two distinct equilibria. However, in molecular systems, we typically set up the system in a non-equilibrium state, and allow it to evolve without further perturbation. We are then generally interested in questions such as: what are the molecular abundances in the eventual equilibrium state, and how fast does the system get there (if at all)? To answer these questions, it is helpful to define {\it biochemical macrostates}.

\begin{figure}
  \includegraphics[width=0.48\textwidth]{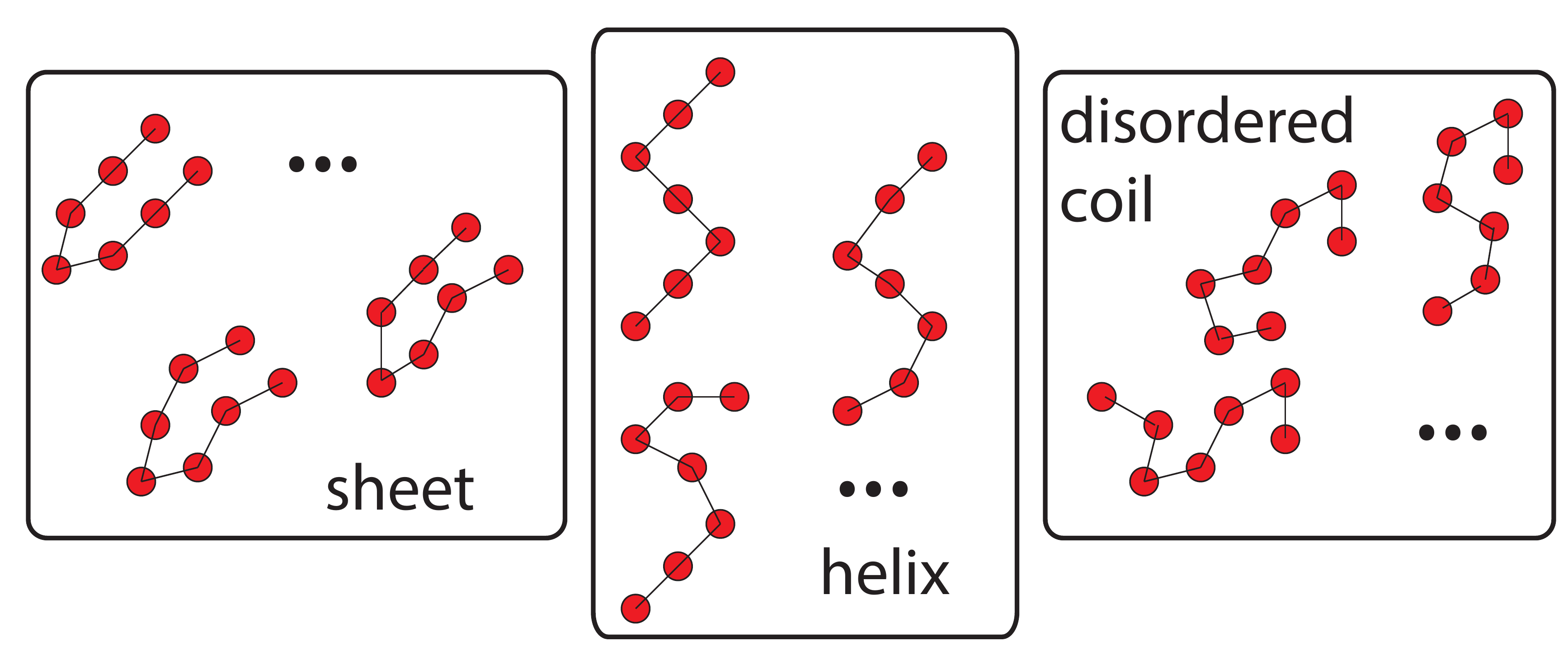}
\caption{Schematic illustration of the division of a continuous set of microstates into a relatively small number of macrostates. In this case there are three discrete macrostates, distinguished by the conformation of a small protein-like molecule. Similar confirmations are grouped together and characteristic examples are shown.  }
\label{fig:macro}       
\end{figure}

\subsection{Macrostates}
\label{sec:macrostates}
The Boltzmann distribution specifies the relative abundances of microstates in equilibrium, and detailed models can be simulated to sample from this distribution. However, microstates are inaccessible in experiment, and inconvenient for theory. Instead, we typically consider biochemical {\it macrostates}. Rather than keep track of all of the atoms in a DNA molecule or protein, we might simply predict theoretically, measure experimentally or infer from simulation the behaviour of the position of the centre of mass, or gross conformational features of a molecular system.  In other words, we group sets of microstates together into a relatively small number of macrostates, as illustrated schematically in Fig.~\ref{fig:macro}. {In this Subsection, we outline the macrostate-level description of thermodynamics that will be used throughout the review.}

In equilibrium, the occupancy of macrostate $i$ is obtained by integrating over all microstates within it
\begin{equation}
P^{\rm eq}_\sigma(i) = \frac{Z_\sigma(i)}{Z_\sigma}= \int_{({\bf x},{\bf p})\in i} {\mathrm d}{\bf x}\,{\mathrm d}{\bf p}\,\rho \frac{\exp(-E_\sigma({\bf x}, {\bf p})/k_{\rm B}T)}{Z_\sigma},
\end{equation}
where we have defined the partial partition function $Z_\sigma(i)$.
{This probability  $P^{\rm eq}_\sigma(i)$ is related to the macrostate free energy $ F_\sigma(i)$ via}
\begin{equation}
 F_\sigma(i) - F^{\rm eq}_\sigma =  -k_{\rm B}T \ln  \left(P^{\rm eq}_\sigma(i)\right) = - k_{\rm B}T \ln \left(Z_\sigma(i)/Z_\sigma \right).
\end{equation}
{Note that} macrostates with high free energy are improbable, and macrostates with low free energy are probable.

In principle, any division into macrostates is valid, although only well-chosen macrostates are helpful. Typically, well-chosen macrostates are either directly identifiable in experiment, amenable to theoretical modelling, or both.  Examples might include macrostates labelled by the number of proteins in dimeric complexes; the end-to-end extension of a biomolecule under stress; or the number of base pairs in a DNA hairpin.

There are two conceptually distinct contributions to the free energy $F_\sigma(i)$ {(Eq~\ref{eq:F})}: the average energy of macrostate $i$,
\begin{equation}
U_\sigma(i) =   \int_{{\bf x},{\bf p} \in i} {\mathrm d}{\bf x},{\mathrm d}{\bf p}\,  E_\sigma({\bf x},{\bf p})P^{\rm eq}_\sigma({\bf x},{\bf p}|i),
\end{equation}
and the entropy of macrostate $i$,
\begin{equation}
S_\sigma(i) = - k_{\rm B}\int_{{\bf x},{\bf p} \in i} {\mathrm d}{\bf x},{\mathrm d}{\bf p}\, P^{\rm eq}_\sigma({\bf x},{\bf p}|i) \ln ( P^{\rm eq}_\sigma({\bf x},{\bf p}|i)/\rho).
\end{equation}
Here $P^{\rm eq}_\sigma({\bf x},{\bf p}|i) = P^{\rm eq}_\sigma({\bf x},{\bf p})/P^{\rm eq}_\sigma(i)$ is the equilibrium probability density of occupying microstate $({\bf x},{\bf p})$ within $i$, given that the system is in one of the microstates within macrostate $i$. A low average energy implies that a macrostate $i$ more probable, since individual microstates with lower energies are more probable. A high entropy $S_\sigma(i)$ implies that many  microstates ${(\bf x}, {\bf p})$ contribute to $i$; for a given average energy, a macrostate with more accessible microstates is more probable.

\section{Free energies of biochemical reactions}
\label{sec:free_energies}
We now discuss the standard statistical mechanical approach to biochemical {\em reactions}. {Our discussion will justify the form of chemical potentials in dilute solution, and illustrate the meaning of free energies and standard free energies of reaction.}  In the subsequent sections, this basic framework will be applied to a range of contexts of relevance to natural and engineered molecular systems. Before proceeding, it is worth noting that, as in Section~\ref{sec:partition_fn}, we consider a system $\sigma$ that can exchange heat with its environment $\Sigma$, but which we have implicitly assumed to occupy a fixed volume $V_\sigma$. In chemical contexts, it is often more  natural  to consider a system maintained at constant pressure $p$. In this case, $V_\sigma$ has to shrink or grow in response to reactions that tend to decrease or increase the internal pressure, respectively. In this case, the Gibbs free energy {$G^{\rm eq}_\sigma=F^{\rm eq}_\sigma+PV^{\rm eq}_\sigma$}, which plays a similar role to the Helmholtz free energy $F$ in a constant pressure setting, is the key quantity.

Biomolecular processes, however, occur in aqueous solution, and the enormous numbers of water molecules present dominate the pressure exerted by the system \cite{Nelson2004}. Reactions between the relatively small number of solute molecules therefore have almost no effect on the pressure and both  theoretical work and  experimental analyses usually assume constant volume (which is much easier to work with). Nonetheless, free energies of solute states are typically quoted in terms of the Gibbs free energies $G$, and the enthalpy $H$ replaces the average internal energy $U$. For internal consistency, we will continue to use $F$ and $U$, but readers familiar with $G$ and $H$ should treat them as essentially equivalent. It is also worth noting that in biochemistry, it is more common to use the molar gas constant $R$ rather than Boltzmann's constant $k_{\rm B}$. This is simply a question of measurement units; if $R$ is used, all entropies and energies must be given per mole of substance, rather than per particle. 

The starting point for our analysis is to treat the solvent implicitly. Formally, this corresponds to integrating over the solvent degrees of freedom in the partition function (Eq.~\ref{eq:pn_fn}), leaving only effective interactions between solute degrees of freedom. In practice, we often just assume that this can be done, and take the effective solute interactions as an input. We then assume that all solutes, or complexes of solutes, can be assigned to a discrete set of molecular species. These species might include ATP, ADP and inorganic phosphate, or a set of individual DNA strands and their complexes: for example, DNA strand $A$, DNA strand $B$ and duplex $AB$. It is helpful to define  macrostates $\{N\}$  of the entire solution in terms of the abundances of each of these species, $\{N\} = (N_A, N_B ...)$. A typical macrostate of a small system, at this level of description, is schematically illustrated in Fig.~\ref{fig:trimer}.

For each species $j$, we can define the single-molecule partition function in the volume $V_\sigma$, $z_\sigma^j$. This quantity is analogous to the partition function in Eq.~\ref{eq:pn_fn}, but the integral is performed only over the degrees of freedom of a single solute molecule of species $j$ in a volume $V_\sigma$, with the solvent again treated implicitly. It should be noted that $z_\sigma^j$ is generally strongly temperature-dependent. In a dilute solution, the overall partition function of a macrostate  $\{N\}$ is essentially given by a product of the individual partition functions, since interactions between molecules that are not in a complex are weak. Hence the degrees of freedom for separate complexes are essentially independent and the partition function factorises. Thus
\begin{equation}
Z_\sigma(\{N\}) = \prod_j \frac{(z_{\sigma}^j)^{N_j}}{N_j!}, 
\end{equation}
where the product $j$ runs over all species types, including complexes. The extra factorial term corrects for overcounting of  states that should actually be viewed as indistinguishable, because they are related by the exchange of identical molecules \cite{Frenkel2001,Tuckerman2010,Huang1987}. From this partition function, the free energies of chemical macrostates follow
\begin{align}
F_\sigma(\{N\}) &= -k_{\rm B}T \ln Z_\sigma(\{N\}) \nonumber \\
&= \sum_j F_\sigma^j(N_j) \nonumber \\
&=  -k_{\rm B} T  \sum_j  \ln \left(\frac{({z_\sigma^j})^{N_j}}{N_j!} \right)  \nonumber \\
& \approx  -k_{\rm B} T  \sum_j \left(N_j \ln {z_\sigma^j} - N_j \ln N_j  + N_j \right).
\end{align}
Here, we have highlighted the fact that the free energy decomposes into a sum over the contributions from each species, $F^j_\sigma(N_j)$, which is a result of the diluteness approximation. The final line uses Stirling's approximation of $\ln N! \approx N \ln N -N$, which is highly accurate for large $N$. We can thus easily calculate the chemical potential $\mu^j_\sigma$, which is the increase in system free energy due to the addition of another molecule of species $j$:
\begin{equation}
\mu^j_\sigma =\frac{\partial F_\sigma(\{N\})}{\partial N_j} =  -k_{\rm B}T \ln (z_\sigma^j) + k_{\rm B}T \ln  (N_j),
\end{equation}
In the limit that $N_j$ is large, $\mu^j_\sigma$ is simply the difference in  free energy arising from adding one molecule of species $j$ to the system. 

Since each $z_\sigma^j$ is a partition function for a single molecule of species $j$ in volume $V_\sigma$, it will grow proportionally to $V_\sigma$; doubling the volume doubles the number of accessible positions for the molecule, but changes nothing else. It is thus convenient to normalise using a standard volume $V_0$, typically taken as the volume in which a single molecule would constitute a concentration $\mathcal{C}_0 =1/V_0$ of 1 mole per litre. Thus
\begin{equation}
\mu^j_\sigma=  -k_{\rm B}T \ln \left(z_0^j \right) + k_{\rm B}T \ln \frac{\mathcal{C}_j}{\mathcal{C}_0},
\label{eq:mu}
\end{equation}
with a $z_0^j = z_\sigma^j {V_0}/{V_\sigma}$ dependent on the choice of standard volume $V_0$, rather than  system volume $V_\sigma$.
This decomposition separates the chemical potential $\mu_j$ (or free energy per molecule of species $j$) into a  term that depends only on the details of the effective interactions within the species,  and a concentration-dependent term \cite{Nelson2004,Huang1987}.  

The chemical potential appearing in Eq.~\ref{eq:mu} is of enormous use in analysing the thermodynamics of molecular systems. In particular, it enables us to  calculate the difference in free energy between initial and final macrostates after any given molecular reaction, telling us how much more likely the final state is than the initial state in equilibrium. To see why, note that every possible reaction $k$ is associated with a {\emph{stoichiometric vector}} $\nu_{kj}$,  the number of molecules of species $j$ produced by the reaction $k$ (this number is negative if species $j$ is consumed by the reaction).  For example, if  reaction $k$ is $A+B \rightarrow AB$, $\nu_{kA} = -1$, $\nu_{kB} = -1$ and $\nu_{kAB} = 1$. Thus the difference in macrostate free energy due to reaction $k$ is simply \cite{Nelson2004}
\begin{equation}
\Delta_k F_{\sigma} \approx \sum_j \nu_{kj}  \frac{\partial F_\sigma(\{N\})}{\partial N_j}  = \sum_j \nu_{kj} \mu^j_\sigma,
\label{eq:dkF}
\end{equation}
where the approximation is highly accurate if $N_j$ is large. Note that $\Delta_k F_{\sigma} $, like $z_\sigma^j$ and $z_0^j$, is generally strongly temperature-dependent. We  will continue to specify the reaction in question by a subscript on $\Delta$, so that $\Delta_k$ indicates the change due to reaction $k$.

Equally, $\Delta_k F_\sigma $, through $\mu^j_\sigma$, is dependent on the concentrations of the reactants and products. It is common to consider the value of $\Delta_k F_\sigma$ at the reference concentration $\mathcal{C}_j=\mathcal{C}_0=$1\,M, $\Delta_k F^0$,  as a ``standard" free energy of the reaction.  $\Delta_k F^0$ can also be further subdivided into standard energetic and entropic contributions,  $\Delta_k U^0$ and $\Delta_k S^0$. This is convenient for book-keeping purposes, but it should be noted that this ``standard" value is 
 dependent on the (arbitrary) choice of $\mathcal{C}_0$, unless the number of reactants and products is the same. For example, for the simple bimolecular binding reaction $A+B \rightarrow AB$,
\begin{equation}
\Delta_{A+B} F^0 = -k_{\rm B} T \ln \frac{z^{AB}_0}{z^A_0 z^B_0}, 
\label{eq:dF0}
\end{equation}
and the term inside the logarithm scales as $V_0$. In such cases, it is inadvisable to over-interpret the sign and magnitude of the standard free energy; whether it is positive or negative depends upon the arbitrary standard concentration, which is 1\,M in general. This dependence on $V_0$ feeds through to the standard entropy $\Delta_k S^0$, so it is also unwise to read too much into the sign and magnitude of this quantity when the numbers of reactants and products differ. Indeed, for  macromolecules such as DNA, RNA and proteins, 1\,M corresponds to an incredibly concentrated solution where the dilute approximations above break down, and behaviour search as the formation of liquid crystals is observed \cite{Nakata2007,DeMichele2012}. For processes involving unequal numbers of reactants and products, therefore, the standard free energy and entropy  exist purely as book-keeping devices, and never describe the actual properties of a reaction. In typical dilute molecular systems, the concentration of the relevant components is orders of magnitude lower. Thus actual values of $\Delta_k F_\sigma$ and $\Delta_k S_\sigma$ are significantly more positive than the standard values for assembly reactions in which the number of reactants exceed the number of products. 

\section{The application of equilibrium thermodynamics to the design of self-assembling systems}
\label{sec:self_assembly1}
\subsection{Self-assembly}
\label{sec:self-assembly}
\begin{figure}
  \includegraphics[width=0.48\textwidth]{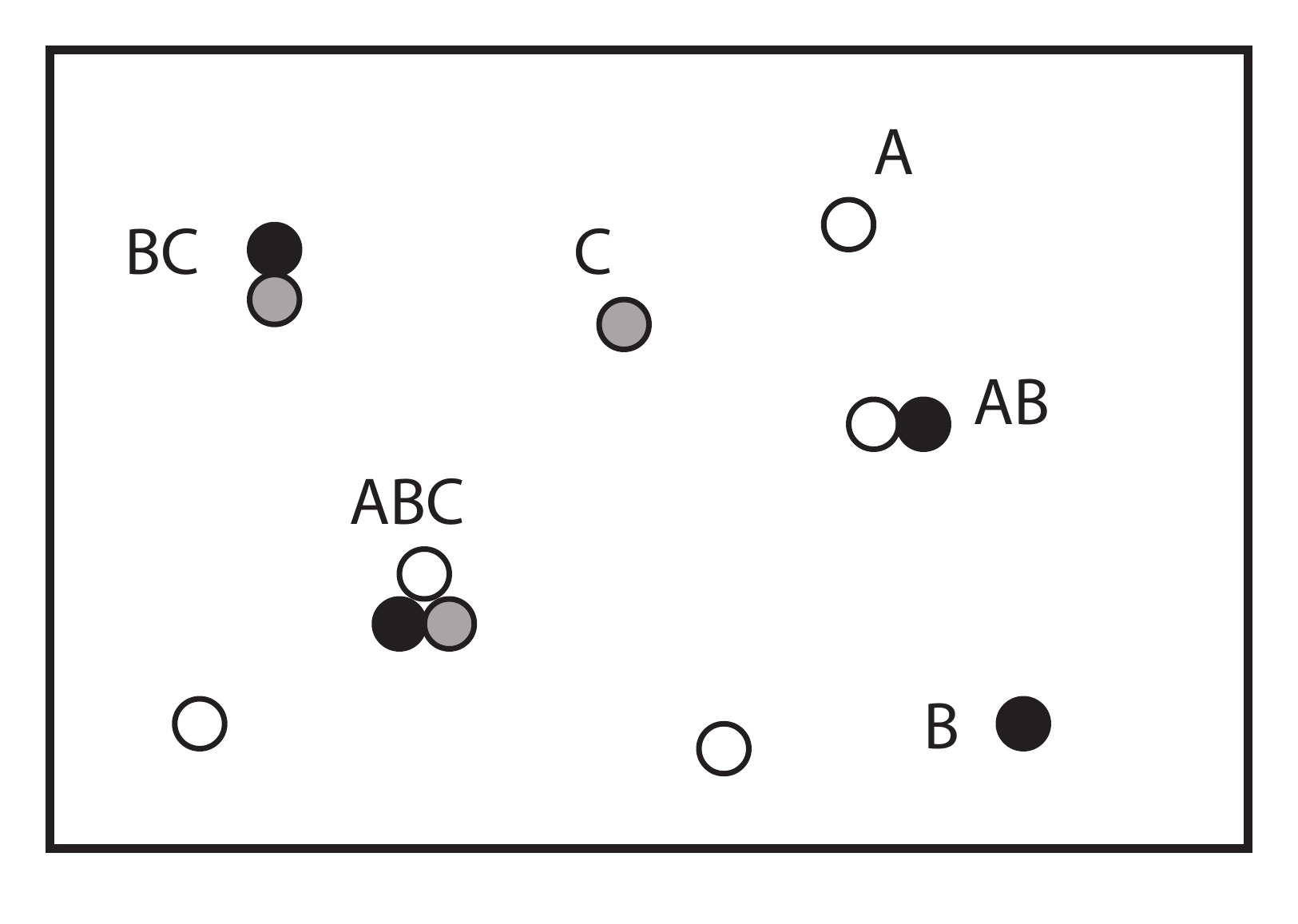}
\caption{A self-assembling system, in which monomers $A$, $B$ and $C$ combine to form a trimer $ABC$. The macrostate in this snapshot is $N_A=3$, $N_B=1$, $N_C=1$, $N_{AB}=1$, $N_{BC}=1$, $N_{AC}=0$ and $N_{ABC=1}$. }
\label{fig:trimer}       
\end{figure}

Self-assembly occurs when molecules are mixed and autonomously bind to produce non-trivial structures. It should be distinguished from step-by-step directed synthesis in which each stage is separately coordinated by an experimenter through careful manipulation of solution conditions \cite{Chen91}, 
 although temperature ramps are often used to optimise results, as analysed in some detail in Ref. \cite{Sobczak2012}. In biology, functional protein complexes \cite{Ahnert2015} and virus capsids \cite{FraenkelConrat55,Johnston2010} must assemble accurately from their components; the rise of nucleic acid nanotechnology has facillitated the design of artificial systems that 
can mimic this behaviour, allowing precisely-controlled finite-size nanostructures \cite{Rothemund06,Ke2012,Goodman2005,Tikhomorov2017}. {In this Section, we outline how the basic thermodynamics introduced in Sections~\ref{sec:fundamentals} and \ref{sec:free_energies} both shape the fundamental behaviour of self-assembling systems,  and guide the details of system design through tools such as Nupack \cite{Dirks2007}.}
To provide context, we will apply the general results to the challenge of assembling three molecular building blocks $A$, $B$ and $C$ into the complex $ABC$, as illustrated in Fig.~\ref{fig:trimer}. Molecules $A$, $B$ and $C$ could each be DNA strands or proteins, for example.  

The simplest design strategy for self-assembly is to ensure that microstates with many well-assembled structures are common in equilibrium. The most likely macrostate $\{N\}$ is the one that maximises $Z_\sigma(\{N\})$, or equivalently minimises $F_\sigma(\{N\}) $,  subject to the constraints of stoichiometry. If  {a large number of molecules are present}, as in typical experiments, fluctuations about the most likely macrostate are relatively small in equilibrium \cite{Frenkel2001,Tuckerman2010,Huang1987,Jaynes1957}, and hence we can infer equilibrium properties purely by analysing this most likely macrostate. 

\subsection{Identifying the typical behaviour of a self-assembling system in equilibrium}
{In this Subsection, we {introduce the basic \emph{mass-action equilibria} underlying the equilibrium yield of self-assembling systems, and demonstrate how they} arise from identifying the microstates with low $F_\sigma(\{N\})$ in dilute solutions. We then discuss the key properties of the resultant equilibria.} 
To minimise $F_\sigma(\{N\})$, we must find $\{N\}$ such that no possible change of $\{N\}$ due to a reaction could reduce $F_\sigma(\{N\})$. In the limit of many molecules, this task is equivalent to identifying the $\{N\}$ for which $ F_\sigma(\{N\}) - F_\sigma(\{N^\prime\})=0$ for every possible reaction $\{N\} \rightarrow \{N^\prime\}$.
From Eq.~\ref{eq:dkF} we therefore require
\begin{equation}
\Delta_k F_{\sigma} = \sum_j \nu_{kj} \mu^j_\sigma =0,
\label{eq:eqm_nu_mu}
\end{equation}
Thus the typical equilibrium state is  the one in which chemical potentials are balanced  for all reactions -- a widely exploited result \cite{Nelson2004,Huang1987}.  

Substituting our expression for $\mu_\sigma^j$ (Eq.~\ref{eq:mu}) into Eq.~\ref{eq:eqm_nu_mu}, we immediately see that at equilibrium
\begin{equation}
 \prod_j \left(\frac{\mathcal{C}_j} {\mathcal{C}_0} \right)^{\nu_{kj}} = \prod_j \left(z_0^j \right)^{\nu_{kj}},
\end{equation}
for all reactions $k$. In the case of  $A+B \rightarrow AB$,
\begin{equation}
 \frac{\mathcal{C}_{AB}}{\mathcal{C}_A \mathcal{C}_B}= \frac{1}{\mathcal{C}_0} \frac{z^{AB}_0}{z^A_0 z^B_0} = K_{A+B}^{\rm eq},
\label{eq:Keq}
\end{equation}
in which we have introduced the equilibrium constant  $K_{A+B}^{\rm eq}$. The quantity $K_{A+B}^{\rm eq}$ is known as a constant because it depends only on the details of the interactions within each species as represented through $z_0^j$; it is independent of system volume of the number of molecules present, although it will depend on quantities such as the temperature. It is also independent of the arbitrary reference volume $V_0$, since each $z^j_0$ scales with this volume. The result is ubiquitous in physical chemistry, and immediately generalises for other reactions  \cite{Nelson2004}. 
 
 If $ K^{\rm eq}$ is known  from earlier experiments for each reaction $k$, or can be predicted from underlying theory, then variants of Eq.~\ref{eq:Keq} can be constructed for each possible reaction and the typical concentrations in equilibrium can be found  by solving the resultant simultaneous equations (when augmented with any conservation laws). Note that a relation such as Eq.~\ref{eq:Keq} exists for every reaction at equilibrium, regardless of whether the reactants are involved in other reactions. Additionally, the same procedure can be followed for not just a single reaction $k$, but a series of reactions. For example, the combined reactions $A+B \rightarrow AB$ and $AB + C \rightarrow ABC$ imply that the relationship  
\begin{equation}
 \frac{\mathcal{C}_{ABC}}{\mathcal{C}_{A}{\mathcal{C}_B}\mathcal{C}_C}= \frac{1}{\mathcal{C}_0^2} \frac{z^{ABC}_0}{z_0^A z_0^B z_0^C } =K_{A+B+C}^{\rm eq},
\label{eq:Keq_ABC}
\end{equation}
is meaningful even if there is no direct $A+B+C \rightarrow ABC$ reaction (which would be essentially impossible without the formation of intermediate complexes). 

To design a system that can self-assemble efficiently into $ABC$, therefore, we should choose molecules that provide a large $K_{A+B+C}^{\rm eq}$, and also large $K_{A+BC}^{\rm eq}$, $K_{AB+C}^{\rm eq}$ and $K_{AC+B}^{\rm eq}$, since $ABC$ needs to out-compete other potential species. Similarly, low values of $K^{\rm eq}$ for reactions that produce off-target species such as $AABB$ are important in preventing mis-assembly. 

An often under-appreciated fact is that {reactant concentrations are important in determining yields}; it is not uncommon to hear a ``melting temperature"  -- approximately the point at which 50\% of the maximum possible yield has been reached -- quoted without reference to component concentrations. However, the fractional yield in equilibrium is sensitive to the initial concentrations of reactants. Lower initial concentrations of $A$, $B$ and $C$ imply a larger reduction in the concentration of complexes in equilibrium, since the numerator in relationships such as Eq.~\ref{eq:Keq_ABC} must change enough to compensate for the reduction of all of the concentrations in the denominator.

\begin{figure}
  \includegraphics[angle = -90, width=0.48\textwidth]{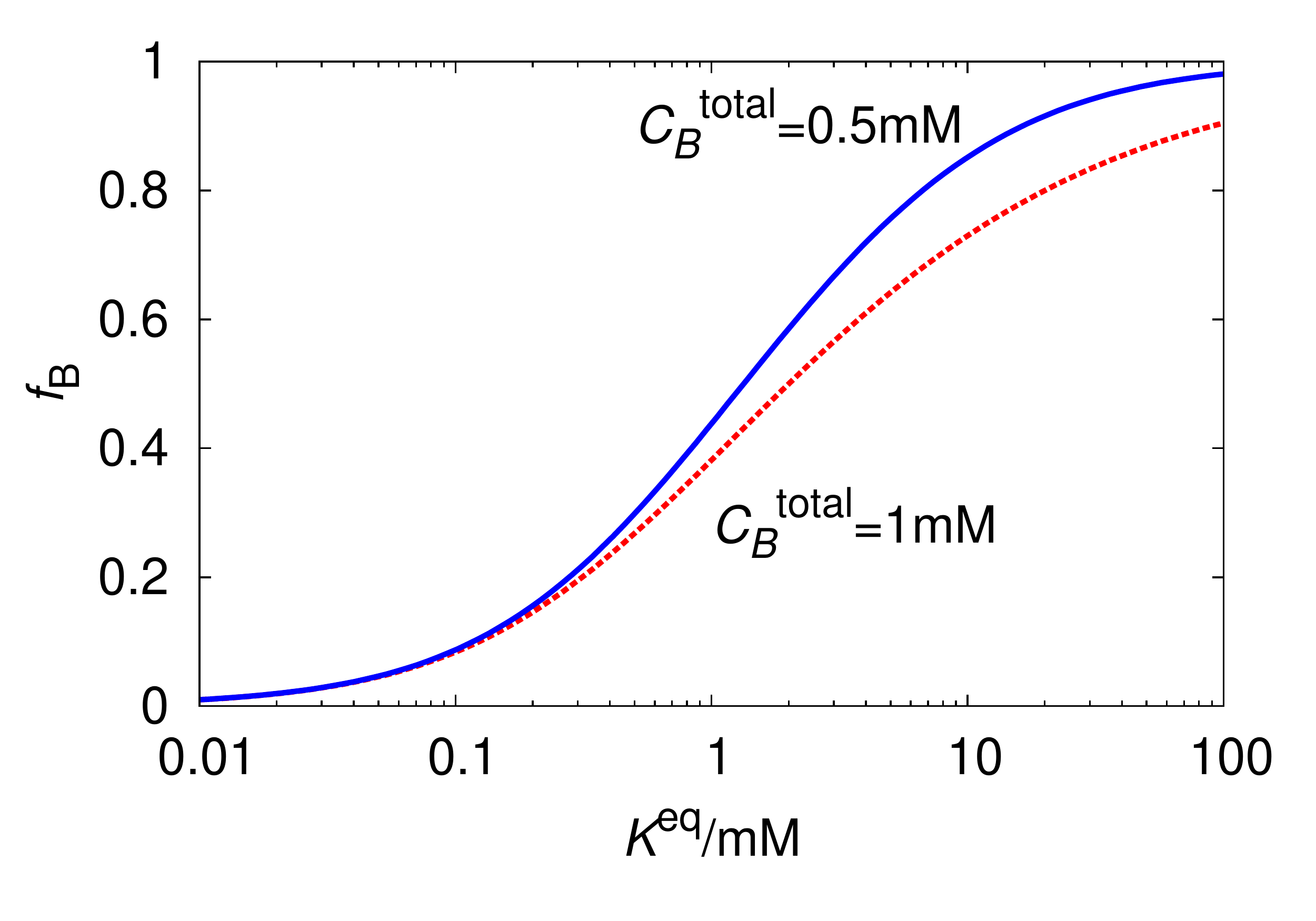}
\caption{Non-stoichiometric conditions can increase the fractional of certain molecules incorporated into target complexes. We plot the equilibrium fraction of $B$ molecules incorporated into $AB$ ($f_B$) by the reaction $A +B \rightleftharpoons AB$, as a function of $K^{\rm eq}_{A+B}$.  stoichiometric mixture. Yield curves are plotted for $\mathcal{C}_B^{\rm total} = \mathcal{C}_A^{\rm total} = 1$\,mM, and  $A$: $2 \mathcal{C}_B^{\rm total} = \mathcal{C}_A^{\rm total} = 1$\,mM.  }
\label{fig:stoichiometry}       
\end{figure}

 Depending on the context, it may {also} be advantageous to use non-stoichiometric mixtures of components {to increase yields} -- for example, an excess of $A$ and $B$ relative to $C$ when assembling $ABC$. Doing so significantly enhances the fraction of $C$ molecules incorporated into $ABC$ structures in equilibrium, at the expense of leaving a pool of $A$ and $B$ which cannot possibly contribute to a target. To illustrate this effect, we plot the fractional yield  $\mathcal{C}_{AB}/\mathcal{C}_B^{\rm total}$ in the reaction $A+ B \rightleftharpoons AB$ as a function of $K^{\rm eq}_{A+B}$ in Fig.~\ref{fig:stoichiometry}. We consider two cases, one with a stoichiometric mixture $\mathcal{C}_B^{\rm total} = \mathcal{C}_A^{\rm total} = 1$\,mM, and one with an excess of $A$: $2 \mathcal{C}_B^{\rm total} = \mathcal{C}_A^{\rm total} = 1$\,mM. It is clear that the fraction of $B$ molecules incorporated into complexes tends to unity much more quickly in the presence of an excess of $A$. This approach is taken in the construction of scaffolded DNA origami \cite{Rothemund06}; an excess of staples is added to enhance the fraction of scaffold strands incorporated into well-formed structures. {Saturating the system with an excess of one type of strand is  a particularly useful approach if free strands of a complementary sequence are the main potential cause of leak reactions, especially given the possibility of pipetting errors}. Murugan {\it et al.} have further  proposed that concentrations of reactants could be judiciously chosen to avoid the formation of undesired off-pathway structures \cite{Murugan2015}.

Finally, it is worth noting that the equilibrium constants $K^{\rm eq}$ are determined exclusively by the properties of the reactant and product species, through the ratio of the appropriate partition functions. Thus relationships such as Eqs.~\ref{eq:Keq} and \ref{eq:Keq_ABC} {\em cannot} be altered by other molecules that are not produced or consumed by the process, or by the properties of intermediate complexes. A change in the equilibrium constant, as measured through the relative concentrations in equilibrium, can only arise from a change in the biochemistry of the initial and final species. This fact will be important when we discuss the role of non-equilibrium catalysts in Section~\ref{sec:catalysis}.

\subsection{The meaning of the equilibrium constant, and estimating its value}
\label{sec:K_eq} 
Relationships between concentrations such as Eqs.~\ref{eq:Keq} and \ref{eq:Keq_ABC}  can also be derived by assuming that reactions obey mass-action kinetics: {\it i.e.,} that rates are proportional to 
concentrations of reactants involved. However, the statistical mechanical approach has two important advantages that {are important in the context of self-assembly. We now proceed to outline these advantages}.

Firstly, the fact that relationships such as Eqs.~\ref{eq:Keq} and \ref{eq:Keq_ABC} hold even without a direct one-step reaction linking reactants and products, and regardless of whether the species are involved in other reactions, is made clear from this thermodynamic perspective. But perhaps even more importantly, the derivation presented highlights the physical meaning of the equilibrium constant $K^{\rm eq}$; it is determined by the ratios of partition functions of the molecular species involved in the reaction. 
The quantities $z_0^j$ are partition functions for complex $j$ in volume $V_0$; species with favourable (low) internal energies, or many accessible states, are favoured.

Crucially, these quantities $z_0^j$ (or at least the relevant ratios) can often be predicted by simple theoretical models. For complex models \cite{Orozco2003,Cino2012,Marrink2007,Doye2013}, ratios of partition functions are the natural quantities to extract from simulation \cite{Ouldridge_bulk_2010,Ouldridge_bulk_2012}, enabling direct prediction of equilibrium constants and hence comparison to experiment. Simpler approaches such as the nearest-neighbour models of DNA and RNA thermodynamics use basic postulates about $z_0^j$ to make analytic predictions of the ratios of partition functions, and hence equilibrium constants, of an enormous number of self-assembling systems using a small set of parameters \cite{SantaLucia2004,Turner2010}. These widely-used tools have been a fundamental component of the growth of nucleic acid nanotechnology and have facilitated the analysis of natural RNA circuitry in cells, for example in Ref. \cite{Borujeni2016}. 

It is hard to overestimate the usefulness of such a predictive tool, even given its finite accuracy. Without it, the systematic design of complex nucleic acid circuits from scratch would be far more challenging -- particularly in terms of eliminating unintended interactions -- and would require the measurement of many equilibrium constants $K^{\rm eq}$ for each design. 

\begin{figure}
  \includegraphics[width=0.48\textwidth]{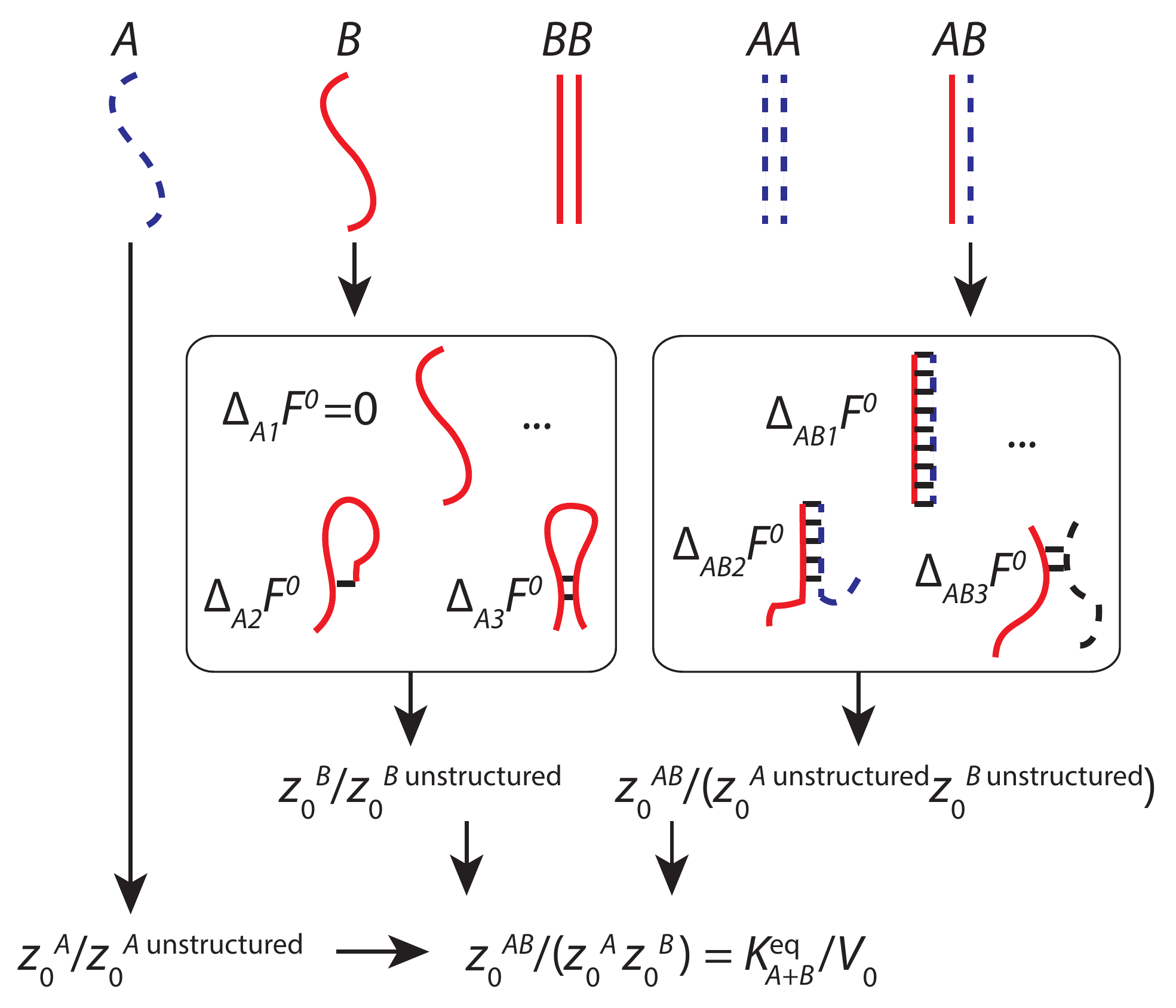}
\caption{A schematic illustration of the calculation of equilibrium constants by algorithms such as Nupack \cite{Dirks2007}, based on nearest-neighbour models of nucleic acid thermodynamics \cite{SantaLucia2004,Turner2010}. Firstly, all possible complexes up to a maximum size (in this case 2) are enumerated. For each of these complexes, and the individual strands, the possible-base pairing macrostates are identified and assigned a free energy relative to the unstructured state. Summing over all contributions for one complex gives the partition function of the complex relative to the unstructured constituent strands; combining this with similar quantities estimated for the single strands themselves gives $K^{\rm eq}_{A+B}$.}
\label{fig:nupack}       
\end{figure}

It is instructive to consider how the nearest-neighbour model can be used by utilities such as Nupack \cite{Dirks2007} to predict the concentration of complexes formed in equilibrium after mixing nucleic acids. Firstly, given a set of strands, all possible complexes below a certain size can be identified (Fig.~\ref{fig:nupack}). For each of these complexes $j$, it is then necessary to  estimate the equilibrium constant $K^{\rm eq}_j$ for complex formation from the constituent single strands at the appropriate temperature -- with these equilibrium constants, the task reduces to solving a set of simultaneous equations involving expressions such as Eqs.~\ref{eq:Keq} and \ref{eq:Keq_ABC} and conservation laws. 

Equilibrium constants are estimated by working at the level of macrostates defined by the pattern of base pairing -- a finer resolution than simply identifying the complexes present, but still far from a true microscopic enumeration of microstates. To estimate  $K^{\rm eq}_{A+B}$, for example, all base-pairing  macrostates of both the complex $AB$ and the individual strands $A$ and $B$ are enumerated (Fig.~\ref{fig:nupack}). For each macrostate $i$, the nearest-neighbour model  predicts the standard free energy relative to a completely unstructured (base-pair free) macrostate using a small set of universal parameters that depend on the sequence of base-pair steps, and the context at the end of the continuous base-paired stations. From these free energies, the relative partition functions of macrostates can be calculated, and by summing over the partition function contributions from all macrostates for both the complex and the individual strands, the ratio $z_0^{AB}/(z_0^A z_0^B )$ can be estimated. Hence $K^{\rm eq}_{A+B}$ can be predicted through Eq.~\ref{eq:Keq}, and used to infer complex concentrations given the total concentrations of all strands.

\section{Thermodynamics as a basis for the design of kinetic models}
\label{sec:kinetics}
\subsection{The importance of kinetics}
Designing an equilibrium state to be consistent with a high yield of a self-assembling structure is the typical approach taken when engineering a self-assembling structure. Indeed, this is the main strategy employed hitherto in the field of nucleic acid nanotechnology when designing structures. However, the existence of a high yield in equilibrium doesn't guarantee successful assembly in finite time -- the system might become trapped in metastable states, and fail to approach the equilibrium yield over a reasonable timescale. To understand why, it is important to develop  kinetic models that can explore dynamical trajectories taken by systems. Additionally, static self-assembled structures are also not the only possible type of molecular system. There has been recent interest in non-equilibrium or dissipative self-assembly \cite{Timonen2013}, in which assembled structures are maintained in a non-equilibrium rather than an equilibrium steady state by a continuous input of energy. Indeed, this is almost a minimal description of a living organism. On a more detailed level, natural molecular circuits generate motion \cite{Nelson2004,Alberts2002}, act as oscillatory clocks \cite{Paijmans2016}, and dynamically sense their environment \cite{Mehta2012,Mehta2016,Govern_PRL_2014,govern2014}; researchers are now designing artificial systems with similar functionality \cite{Zechner2016,Stricker2008,Zhang2011}. In these dynamical systems, reaction kinetics is inherently important.

{ We now give a detailed discussion of the influence of thermodynamics on the kinetics of molecular systems. We start with a discussion of the biochemical master equation as a fundamental description of biochemical kinetics in Section~\ref{sec:stoch_proc}, including pitfalls associated with poorly-chosen macrostates, before moving on to the constraints on the dynamics imposed by thermodynamic considerations  in Section~\ref{sec:detailed_balance}. Finally, we discuss how to build a dynamical model taking these constraints into account in Section~\ref{sec:designing_models}.}

\subsection{Molecular systems as stochastic processes}
\label{sec:stoch_proc}
Statistical mechanics is inherently probabilistic; the Boltzmann distribution (Eq.~\ref{eq:Bmann}) is, after all, a probability distribution for finding the system $\sigma$ in a given microstate. It is therefore natural to describe system dynamics using a stochastic (or random) process \cite{VanKampen2007} over these microstates.  At a given time $t$, the system occupies a microstate $({\bf x}, {\bf p})$ with a probability density $P_\sigma({\bf x}, {\bf p},t)$; system dynamics lead to an evolution of this distribution over time. After a long time, $t \rightarrow \infty$, and in the absence of external driving, the system should relax towards a stationary (time-independent) distribution given by the Boltzmann distribution $P^{\rm eq}_\sigma({\bf x}, {\bf p})$  of Eq.~\ref{eq:Bmann} (assuming the state space is {\em ergodic}, as is typical).

For an ideal memoryless environment, the stochastic process should be Markovian \cite{VanKampen2007}; {\it i.e.,} the future evolution of $P_\sigma({\bf x}, {\bf p},t)$ only depends on the past via the current value of $P_\sigma({\bf x},{\bf p}, t)$. Equivalently, the outcome of leaving a microstate 
$({\bf x},{\bf p})$ is independent of the route by which the system  reached $({\bf x},{\bf p})$. Of course, when analysing molecular systems, we typically work at the level of chemical macrostates. We describe the system through the abundances of species, or perhaps the hydrogen-bonding patterns,   rather than individual positions and momenta. In this case, our state space is discrete and the stochastic process involves a series of transitions by which the system undergoes discrete hops between macrostates.

Generally, it is assumed that the stochastic process is also Markovian at the macrostate level \cite{Seifert2012,Seifert2011,Hill1989}. In other words, transitions from macrostate $i$ to macrostate $j$ occur at a fixed rate $k_{ji}$ (frequently zero), meaning that the time spent in state $i$ prior to a transition is exponentially distributed; the average lifetime is given by $\tau_i = 1/\sum_j  k_{ji}$; and the destination state is independent of the state from which the system entered $i$ and the length of time spent within $i$. Formally, given this assumption, the probability distribution evolves according to the familiar Master equation \cite{VanKampen2007}. This type of model is so universal that it is often not realised just how big an assumption it is to treat coarse-grained dynamics as Markovian. 

\begin{figure}
  \includegraphics[angle = -90, width=0.48\textwidth]{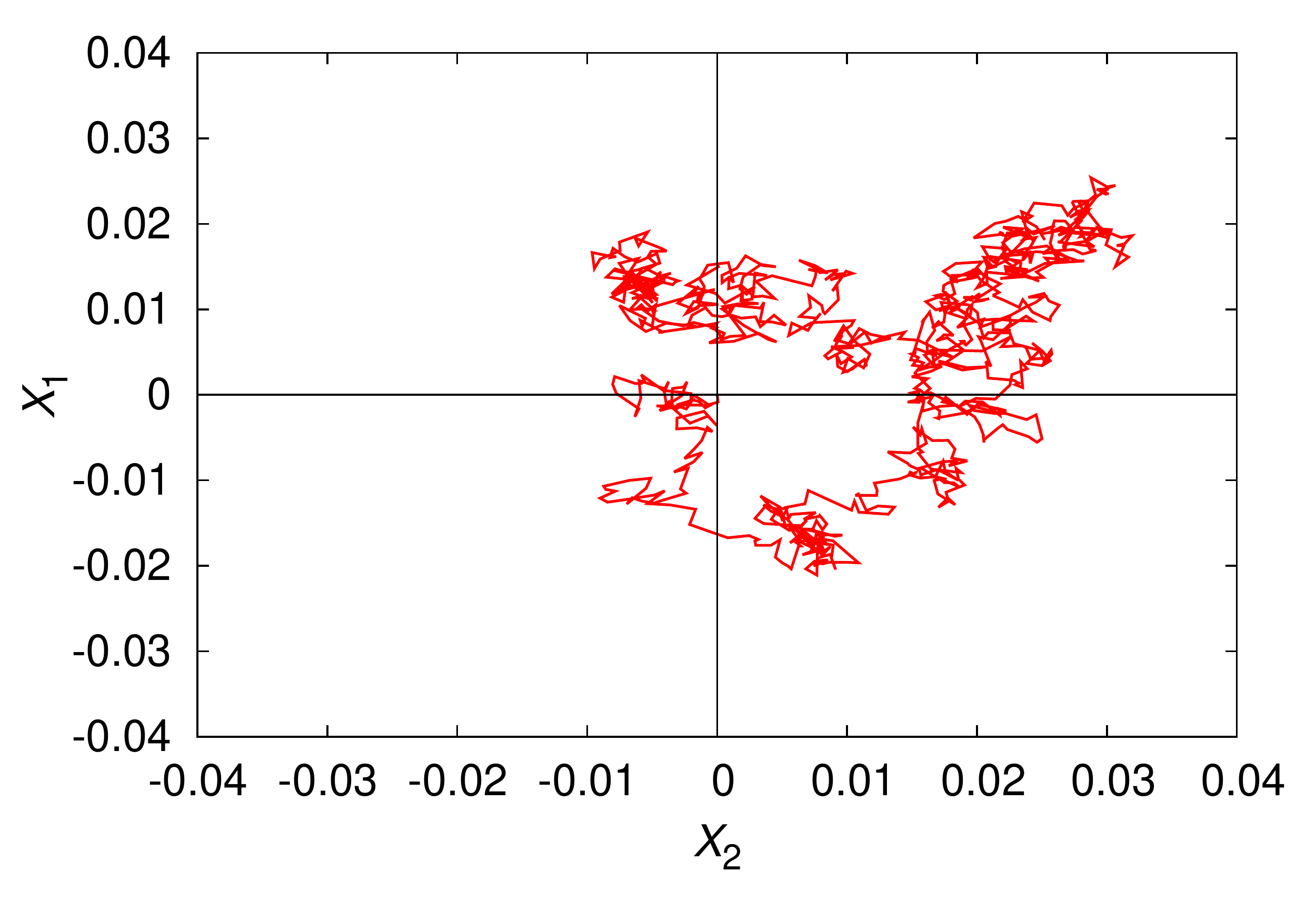}
\caption{Coarse-grained dynamics can be non-Markovian, even if the underlying dynamics at the macrostate level is Markovian. Plotted is the trajectory of a simple Markov process (a 2D Ornstein Uhlenbeck process \cite{VanKampen2007}) over continuous variables $X_1$ and $X_2$. Dividing the state space into macrostates according to the sign of $X_1$ and $X_2$ leads to highly non-Markovian dynamics st the macrostate level; destinations and dwell times do not exhibit the memoryless property. }
\label{fig:markov_nonmarkov}       
\end{figure}

Consider Fig.~\ref{fig:markov_nonmarkov}, in which we have divided a two-dimensional microstate space into macrostates in an essentially arbitrary fashion, and plotted a sample trajectory of a simple process that is Markovian at the microstate level. Fundamentally, even if the dynamics is Markovian at the microstate level, coarse-graining introduces memory into the process. Trajectories that enter macrostate $i$ from macrostate $j$ are close to the border between the two, and hence are likely to quickly cross back  into $j$. As a result, both the transition time {\it and} transition destination should, in general, depend in a complex manner the previous macrostates visited by the system, and the length of time for which the current macrostate has been occupied.

\begin{figure}
  \includegraphics[width=0.48\textwidth]{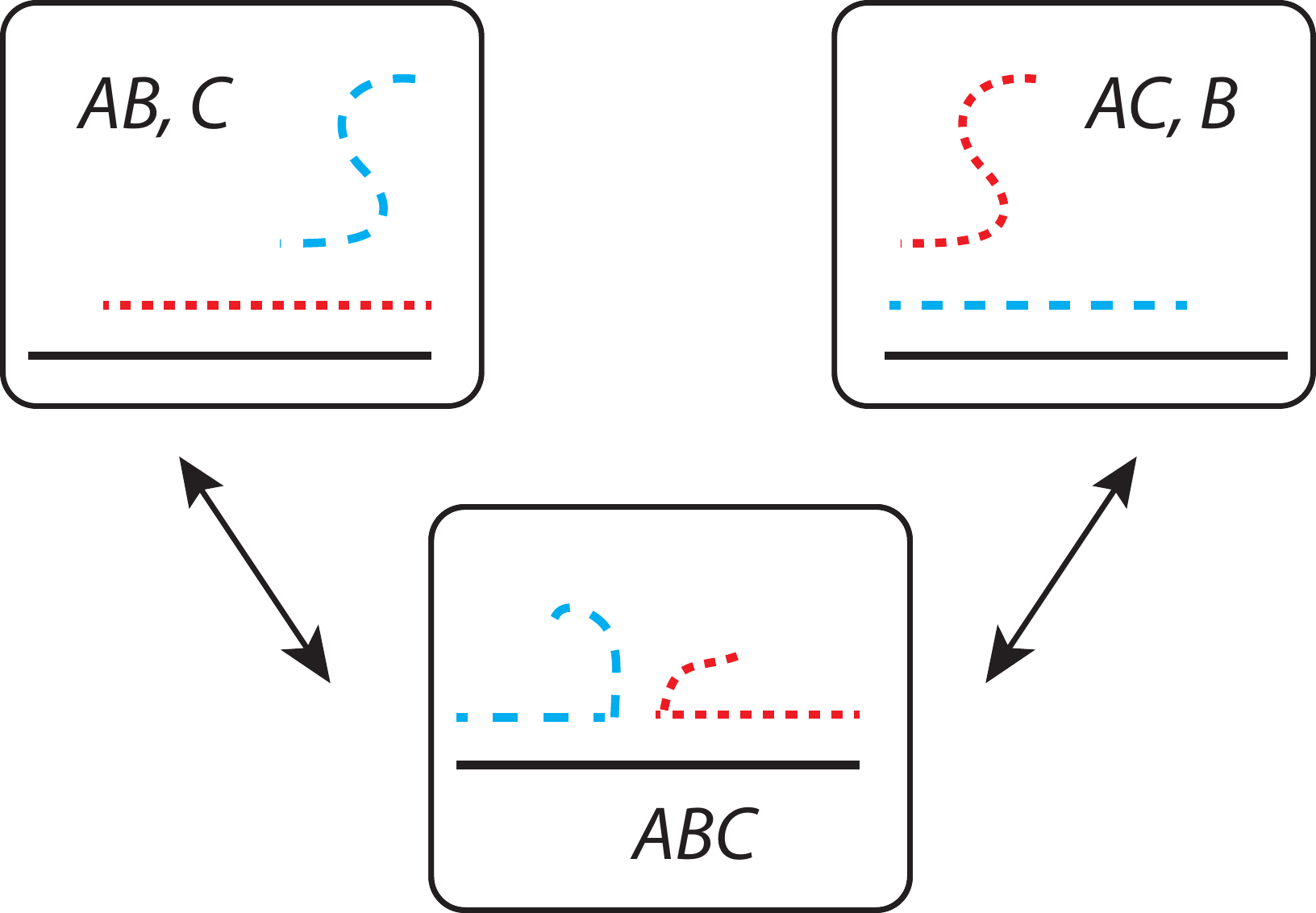}
\caption{A practically-relevant example of a coarse-graining into macrostates that is inappropriate for treatment with Markovian dynamics. Consider a DNA-based toehold-exchange reaction, in which $B$ and $C$ compete for binding to $A$ via strand displacement. It is perfectly possible to define three macrostates as shown above, and indeed doing so would enable the calculation of molecular abundances in equilibrium. However, it is inappropriate to model the dynamics as Markovian, because the microstates of the $ABC$ complex are not typically fully-explored before it dissociates.}
\label{fig:strand_exchange_markov}       
\end{figure}

For a more concrete example, consider the strand exchange reaction shown in Fig.~\ref{fig:strand_exchange_markov}. Strand exchange is a basic process underlying much of DNA computation \cite{Qian2011,Zhang_disp_2009,Chen2013}. It might be tempting to describe the system using three macrostates: $A$ bound to $B$ only; an $ABC$ complex; and $A$ bound to $C$ only. Indeed, such an approach is sensible if one is only interested in the relative abundances of complexes in equilibrium. However, the system dynamics cannot be well-described by a Markov process at this level. The need to initiate and complete branch migration to exchange base pairs between $AB$ and $AC$ duplexes means that an $ABC$ complex formed by $AB + C \rightarrow ABC$  is in reality much more likely to dissociate into $AB + C$ than a complex formed by $AC + B \rightarrow ABC$, violating the assumptions of a Markov process. Even splitting the $ABC$ macrostate into two separate macrostates, depending on whether $AB$ or $AC$ contains the most base pairs \cite{Zhang_disp_2009}, does not provide a satisfactory treatment of the system; it is necessary to resolve macrostates on at least the base-pair level to provide a predictive Markov model of system dynamics \cite{Srinivas2013}. A similar example in cells is the translation of RNA \cite{Reuveni2011}; if ``ribosome bound to RNA" is treated as a single macrostate, rather than modelling codon incorporation as individual steps, a highly unrealistic exponential distribution of times for translation will be obtained.

So when is a Markov assumption reasonable? This is a subtle problem, but the basic idea is that macrostates must be carefully chosen so that a transition involves passing through an unfavourable (high energy or low entropy) set of microstates around the boundary. In this case, the system typically spends a long time fully exploring each macrostate before a sudden hop to a neighbouring one. Transitions, which are complete when the system has fully crossed the unfavourable microstates on the boundary, are rare -- the time taken waiting to see a transition is long compared to the duration of the transition. When these assumptions hold, any memory of the previous macrostate is lost whilst exploring the new one, allowing the Markov assumption \cite{Seifert2012,Seifert2011,Hill1989}. {A more detailed discussion of when coarse-graining is possible, and how to do it systematically, is presented  in Ref. \cite{Bo2017}.}

\begin{figure}
  \includegraphics[angle = -90, width=0.48\textwidth]{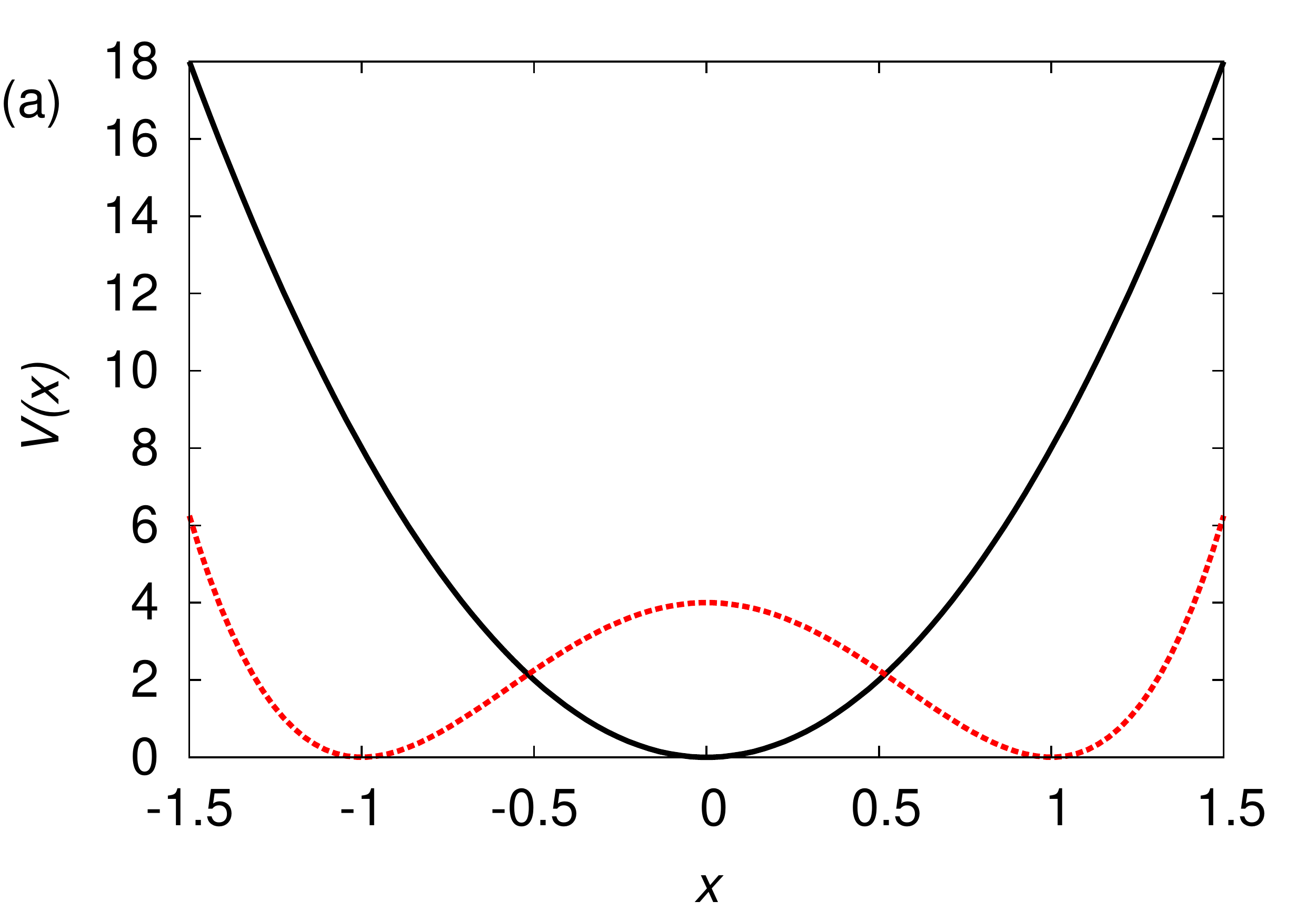}
  \includegraphics[angle = -90, width=0.48\textwidth]{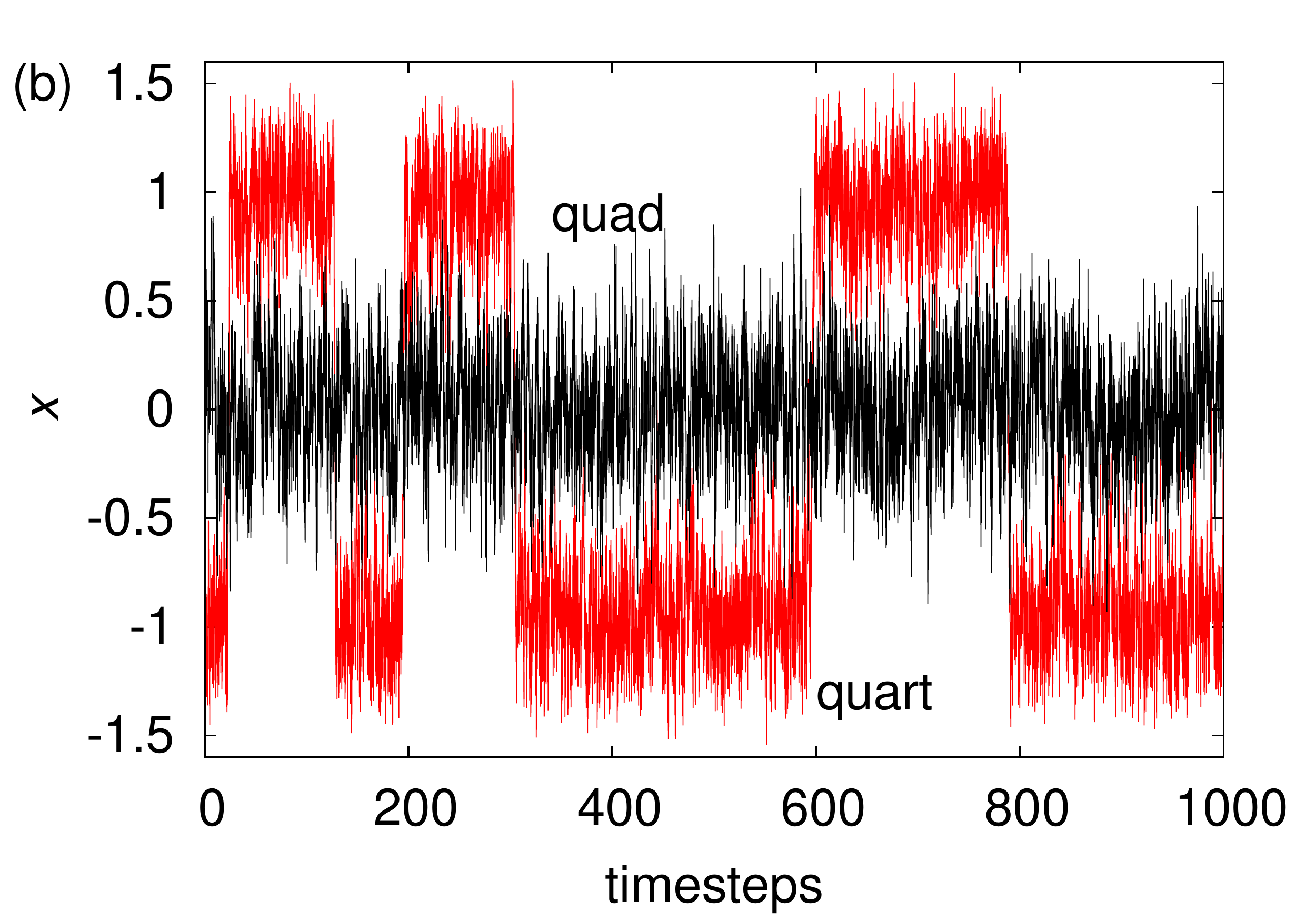}
\caption{Macrostates that are separated by well-defined (free-) energy barriers lead to rare event kinetics and allow for a Markovian treatment. (a) Two potentials: a quadratic well with a single minimum, and a quartic well with two minima separated by a barrier. (b) Example trajectories of (overdamped)  Brownian dynamics in these wells. For the quartic well (trajectory labelled ``quart") transitions are rare events and each well is sampled representatively prior to transitions. It is therefore reasonable to assume Markovian dynamics between macrostates defined by $x>0$ and $x<0$. By contrast, for the quadratic single well (``quad") this is not possible. }
\label{fig:quartic_quadratic}       
\end{figure}

A simple example is given by comparing the trajectories of particles in two potential energy wells. In the first case, the well is quadratic, with a single minimum at $x=0$; in the second, it is quartic, with two minima at $x=-1$ and $x=+1$ (see Fig.~\ref{fig:quartic_quadratic}\,(a)). In both cases it is formally possible to define macrostates according to whether the particle occupies $x<0$ or $x>0$. However, only in the first case, in which the transition from $x<0$ to $x>0$ is associated with climbing over an unfavourable energy barrier, is it reasonable to describe this process using a Markov model at the macrostate level (compare the two trajectories in Fig.~\ref{fig:quartic_quadratic}\,(b)). Returning to the toehold exchange reaction in Fig.~\ref{fig:strand_exchange_markov}, the three-state Markov model fails because a system that enters the $ABC$ macrostate isn't likely to fully explore that macrostate prior to leaving it. Since branch migration is slow \cite{Srinivas2013}, the  strand that has just bound will often detach before all branch migration intermediates have been explored, invalidating the requirements for a Markov model at the level of these macrostates. 

From this point onwards we will assume a biomolecular system with discrete macrostates that have been well-chosen. Thus the dynamics is Markovian and can be well described by a master equation with rate parameters $k_{ji}$  \cite{VanKampen2007}. What we will say will also be applicable to a full description at the level of microstates  $({\bf x}, {\bf p})$. 

\subsection{Detailed balance}
\label{sec:detailed_balance}
 The transition rates $k_{ji}$ between all pairs of macrostates fully define system behaviour, given a particular initial condition.
Knowledge of system thermodynamics ({\it ie.,} the free energy of macrostates $F_\sigma(i)$) doesn't specify kinetics, but it does place strong and important restrictions. Firstly as noted in Section~\ref{sec:stoch_proc}, the system should eventually relax to the equilibrium (Boltzmann) distribution over macrostates. Thus knowledge of the equilibrium distribution constrains the set of rate parameters $\{k_{ji}\}$ -- they must result in the appropriate steady state. 

\begin{figure}
  \includegraphics[width=0.48\textwidth]{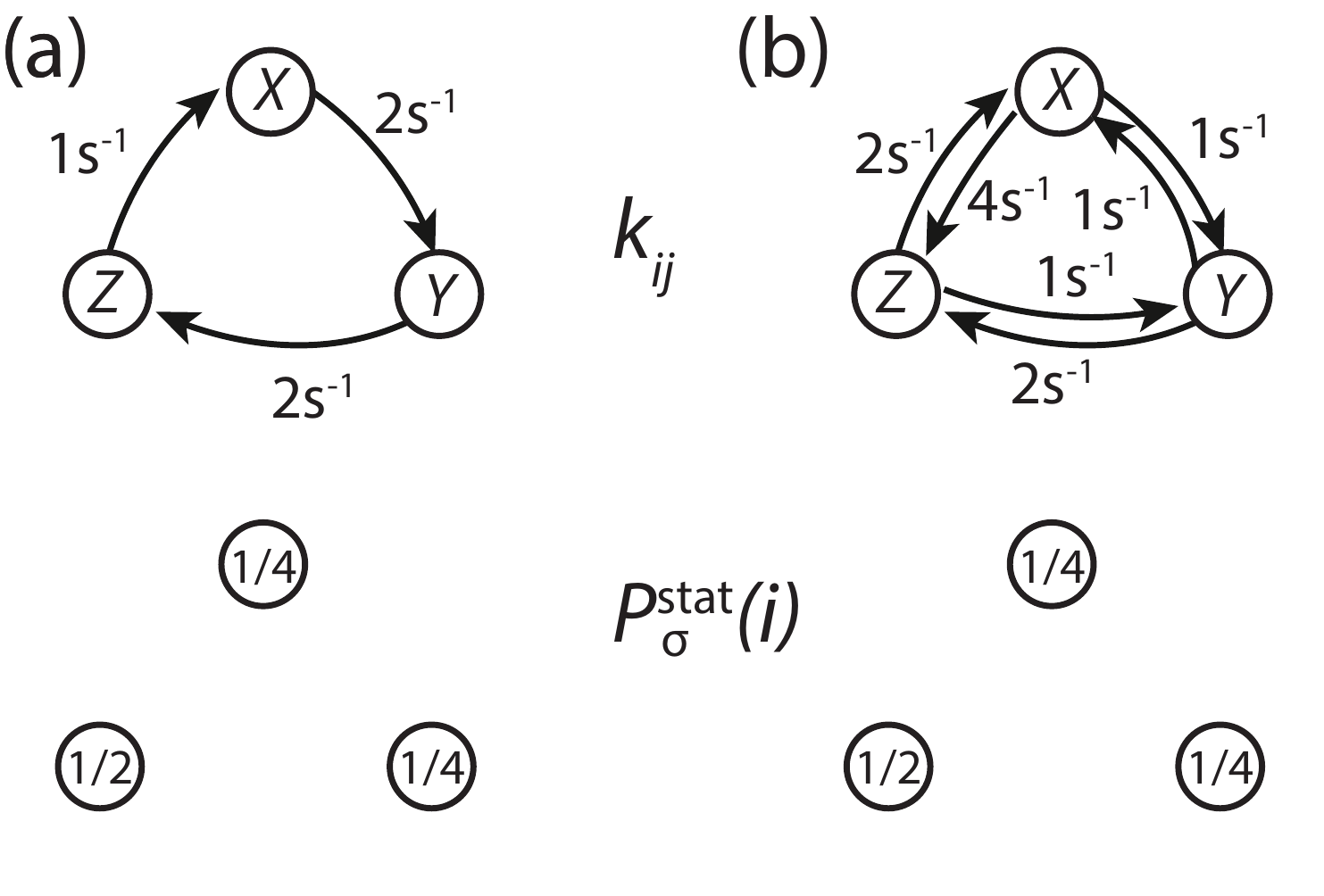}
\caption{Detailed balance in a discrete state Markov process. (a) and (b) represent  Markov processes {defined on} the discrete state space $X,Y,Z$. These processes both have the same stationary distribution $P^{\rm stat}_\sigma(i)$, shown above. However, system (a) exhibits a tendency to flow in a clockwise direction $X \rightarrow Y \rightarrow Z$ even if $P_\sigma(i)= P^{\rm stat}_\sigma(i)$, whereas system (b) exhibits detailed balance; the net number of transitions between each pair of states cancels at $P_\sigma(i)= P^{\rm stat}_\sigma(i) =  P^{\rm eq}_\sigma(i)$, as can be explicitly verified.}
\label{fig:db}       
\end{figure}

Thermodynamic systems, however, are constrained much more tightly than this. A feature of equilibrium is that there should be no tendency of reactions to flow in one direction \cite{VanKampen2007}. {This fact was alluded to when the concept of equilibrium was introduced in Section~\ref{sec:fundamentals}; in equilibrium, there should be no net flux of systems between any pair of microstates}. For example, if a single molecule can occupy three different conformational macrostates, $X$, $Y$ and $Z$, one could imagine a steady state with a systematic flow $X \rightarrow Y \rightarrow Z \rightarrow X$ (Fig.~\ref{fig:db}\,(a)). Such a steady state is {\it impossible} in equilibrium. If it existed, it would be possible to use the systematic flow to power molecular machines (Section~\ref{sec:catalysis}), which would violate the second law of thermodynamics. This feat would be analogous to powering a water mill using a completely flat and undisturbed pond. 

Instead, in equilibrium, each individual transition must be balanced by its microscopic reverse (Fig.~\ref{fig:db}\,(b)); the total rate at which $X \rightarrow Y$ transitions are observed should balance the total rate of $Y \rightarrow X$. This feature is known as the {\em principle of detailed balance}, and it is a central plank of the thermodynamics of molecular systems. In terms of the rate parameters, equating the total number of transitions per unit time $\alpha$ in both directions in  equilibrium gives
\begin{equation}
\alpha_{i \rightarrow j} = k_{ji} P^{\rm eq}_\sigma(i) = k_{ij} P^{\rm eq}_\sigma(j) = \alpha_{j \rightarrow i},
\end{equation}
which clearly holds in Fig.~\ref{fig:db}\,(b), but not Fig.~\ref{fig:db}\,(a). Thus
\begin{equation}
\frac{k_{ji}}{k_{ij}} = \frac{P^{\rm eq}_\sigma(j)}{P^{\rm eq}_\sigma(i)} = \exp(-(F_\sigma(j) - F_\sigma(i))/k_BT)
\label{eq:db}
\end{equation}
for a simple chemical system obeying detailed balance. 
The ratio of rate parameters $k_{ji}/k_{ij}$ is then determined exclusively by the difference in free energies between macrostates $i$ and $j$. It should be immediately evident that rate parameters constrained in this way will necessarily lead to the appropriate equilibrium distribution $P^{\rm eq}_\sigma(i) \propto \exp(-F_\sigma(i)/k_{\rm B}T)$, since  by definition all pairs of transitions will cancel out if $P_\sigma(i) = P^{\rm eq}_\sigma(i)$. Importantly, since the rate parameters are {\em constant}, the relationship between rate constants in Eq.~\ref{eq:db}  is a feature of the dynamics and holds for any $P_\sigma(i)$, even when the system is out of equilibrium and $P_\sigma(i) \neq P_{\sigma}^{\rm eq}(i)$. 

An important consequence of detailed balance is that we can now see which transitions will tend to occur spontaneously in representative trajectories. If $F_\sigma(i) > F_\sigma(j)$, we expect to see systems move from $i$ to $j$ more quickly than they would from $j$ to $i$. In equilibrium, this tendency is compensated for by the relative population size, yielding detailed balance. This is true even if it takes several steps to reach $j$ from $i$, since Eq.~\ref{eq:db} holds for each of those steps. If $(F_\sigma(i) - F_\sigma(j))/k_{\rm B}T \gg 1$, then we expect to see $i \rightarrow j$ occur spontaneously during trajectories, but we will essentially never observe a system starting in $j$ and transitioning to $i$ (unless we force it from the outside). In the context of a chemical reaction, for example $A + B \rightleftharpoons AB$, we expect to see systematically more transitions from left to right whilst $\Delta_{A+B} F_\sigma = \mu_\sigma^{AB} - (\mu_\sigma^A + \mu_\sigma^B)>0$. On a large scale with many molecules, we will effectively see the reaction spontaneously proceed in one direction determined by the sign of $\Delta_{A+B} F_\sigma$, until equilibrium ($\Delta_{A+B} F_\sigma = 0$) is reached.

\subsection{Parameterising a kinetic model for molecular systems}
\label{sec:designing_models}
{
The relative rate of forwards and backwards transitions is thus fixed by the associated change in macrostate free energy. In this subsection, we discuss how this simple self-consistency relation places strong constraints on the rate parameters that can be used to describe a biochemical system, if it is to be thermodynamically well-defined. Doing so is particularly important in developing physically reasonable models of self-assembly, polymerization or depolymerization \cite{Dannenberg2015,Nguyen2016,Andrieux2008,Sartori2015,Gaspard2016}, and also nanotechnologically important reactions such as strand exchange \cite{Zhang_disp_2009}. Unless this physical constraint is applied when parameterising such models, unphysical cyclic flows of reactions will be observed in steady state. Preserving the relationship between free energy change and relative reaction rates is also essential if the costs of fuel-consuming systems \cite{Mehta2012,Lan2012,Barato2015,Pietzonka2016,Ouldridge_copy_2015}, which will be discussed in Section~\ref{sec:catalysis}, are to be understood.}

{Note that this requirement of thermodynamic self-consistency is not unique to a particular approach to modelling a biochemical reaction network. We have been considered a fully-stochastic description at the level of the chemical master equation, \cite{VanKampen2007,Gillespie2009}, but similar reasoning also applies to modelling performed in the deterministic limit \cite{Gillespie2009} or using a chemical Langevin approximation  \cite{VanKampen2007,Gillespie2009}. All of these approaches are potentially thermodynamically well-defined, but care must be taken when parameterising.}

{It is sufficient to consider how transition rates defining the chemical master equation should be chosen, since these directly determine the deterministic and Langevin approximations to the system.} It is typical to assume that in dilute solutions, rates per unit volume are proportional to concentrations of reactants. For example, in the reaction $A+B \rightarrow AB$, the macrostate-dependent rate for the binding transition
 $ k_{N_{AB}+1,N_{AB}}\left( { N}_{A}, {N}_B  \right)  $  is given by
\begin{equation}
 k_{N_{AB}+1,N_{AB}}\left( {N}_{A}, N_B  \right) = \phi_{\rm bind} \frac{N_A}{V_\sigma} \frac{N_B}{V_\sigma} V_\sigma,
\end{equation}
where  $\phi_{\rm bind}$ is a  bimolecular rate constant, and the reverse rate is given by
\begin{equation}
 k_{N_{AB},N_{AB}+1}\left( N_{AB}\right) = \phi_{\rm unbind} \frac{N_{AB}}{V_\sigma}  V_\sigma,
\end{equation}
where $\phi_{\rm unbind}$ is a unimolecular rate constant for unbinding. {Such a choice is potentially thermodynamically consistent, but Eq.~\ref{eq:db}
combined with Eqs.~\ref{eq:mu} and \ref{eq:dF0} implies a specific relationship between the two rate constants. The familiar result is:}
\begin{equation}
\frac{\phi_{\rm bind}}{\phi_{\rm unbind}} = V_0 \exp(-\Delta_{A+B} F^0/k_{\rm B}T).
\label{eq:rate_const_ratio}
\end{equation}
We have thus seen how a constraint on the transition rate parameters in the chemical master equation translates into constraints on the familiar first and second order rate constants of association and dissociation. Similar results can be obtained for other reactions.

{When constructing a kinetic model, it is advisable to start from a model for the free energies $F_\sigma(i)$, and then impose Eq.~\ref{eq:db} or  Eq.~\ref{eq:rate_const_ratio} for each pair of reactions. Indeed, it is }extremely hard to write down directly a set of transition rates $\{ { k} \}$ or rate constants $\{{ \phi}\}$ that respect a sensible free energy model for a complex system. Typically, the free energy difference implied by two distinct pathways between microstates will be inconsistent, resulting in unwanted steady-state reaction fluxes {\it etc.}

As an example, we might consider a small DNA nanostructure, whose assembly involves binding and unbinding of duplex sections. A simple assumption might be that all binding transitions have the same bimolecular rate constant  (perhaps $\sim 10^6$\,M\,s$^{-1}$), which sets $k_{ij}$ for all binding transitions. The inverse $k_{ji}$ transition rates then follow from the free energy of binding, $F_\sigma(i) - F_\sigma(j)$, which might be estimated via the nearest-neighbour model.

In general, both the problem of estimating $F_\sigma(i)$ and the absolute rates of one of each pair of transitions can be subtle \cite{Srinivas2013,Flamm2000,Schaeffer2015,Dannenberg2015}, and the consequences for the dynamics can be profound. In particular, it is not always obvious how a change in $\Delta_{k} F_\sigma$ might be manifest in the rates. For example, consider the association of two DNA strands $A$ and $B$ that can form hairpins in the single-stranded state. These hairpins serve to make the standard free energy of formation of a duplex  $\Delta_{A+B} F^0$ less favourable (less negative), and hence reduce the ratio $\phi_{\rm bind}/\phi_{\rm unbind}$ through Eq.~\ref{eq:rate_const_ratio}. It might be natural to assume that hairpins essentially reduce the rate of binding $\phi_{\rm bind}$, and indeed hairpins are deliberately used for this purpose to create metastable systems \cite{Turberfield2003,Yin2008,Choi2014,Meng2016}.  However, experimental evidence suggests that in some cases,
$\phi_{\rm bind}$ is reduced by much less than $\exp(-\Delta_{A+B} F^0 /k_{\rm B}T)$, and implying that in fact most of the reduction in free energy is manifest as an increase in the off-rate $\phi_{\rm unbind}$ \cite{Gao2006}. This observation is supported by detailed simulation \cite{Schreck2015}, in which it is observed that hairpins form prior to full dissociation and stabilise the partially-melted state, accelerating unbinding. Similarly, experimental  work supported by theory shows that the overall forwards rate of strand displacement and exchange reactions can be adjusted by orders of magnitude at a fixed overall standard free energy of reaction \cite{Zhang_disp_2009,Machinek2014}.

Despite these subtleties, simple kinetic models of complex molecular systems can provide deep insight into function. For example, the exquisite data provided by Zhang and Winfree on the rates of DNA strand displacement and toehold exchange reactions \cite{Zhang_disp_2009} allow thermodynamically consistent modelling of those key processes, underlying the systematic design of complex molecular circuits using tools such as DSD \cite{Lakin2011}. In turn, the physical basis of these parameters have been probed by more detailed modelling \cite{Srinivas2013} at the base-pair level of description. Similarly, it has long been known that kinetics, rooted in a free-energy landscape, is fundamental to understanding how proteins fold \cite{Dill2012}. Recent work has explored and manipulated the kinetics of assembly DNA origami and DNA brick assembly in a similar fashion, highlighting the importance of subtle cooperative effects \cite{Dannenberg2015,Dunn2015,Reinhardt2014,Song2012}.

\section{{The consequences of reversibility in biochemical reactions, and the relation to thermodynamic reversibility}}
{Reversibility is an important term in both thermodynamics (reversible processes generate no entropy) and the literature on chemical reaction networks (reversible transitions can occur backwards or forwards). In this section, we first explain why the microscopic reverse of all observed reactions must be possible, and  how reaction networks modelled with one-directional transitions should be understood. We then demonstrate the consequences of  microscopic reversibility for a particular molecular computation algorithm, highlighting potential issues with neglecting microscopic reversibility in system design. Finally, we contrast the meaning of  thermodynamic reversibility with that of microscopically reversible transitions in an attempt to clarify a frequently misunderstood dichotomy. }
\subsection{Including reverse transitions in modelling}
\label{sec:reversibility}
One immediate consequence of Eq.~\ref{eq:db} is that if a forwards reaction is possible, so too is its reverse, with relative rates determined by the initial and final free energies. This concept is known as {\it the principle of microscopic reversibility}, and has far-reaching consequences. This principle suggests that both forwards and backwards transitions should always be explicitly included in any model.  However, successful modelling is often done by treating certain reactions as totally irreversible. For example, nobody models transcription of RNA in cells by considering the reverse process by which RNA returns to the DNA and is destroyed base by base whilst in contact the gene that encoded it \cite{Bennett1982}. Similarly, strand displacement kinetics is often fitted by assuming that the reaction proceeds to 100\% completion \cite{Machinek2014}.

In general, whether it is reasonable to neglect reverse transitions depends upon the context, and the purpose of the modelling. Processes such as RNA transcription involve constant input of chemical fuel (this situation will be discussed in Section~\ref{sec:catalysis}), and are thus far from equilibrium; in such cases, backwards transitions can be rendered irrelevant by the presence of alternative pathways. For example, RNA is digested by exo- and endonucleases in the cell, instead of needing to be destroyed by the reverse of transcription. Unless the modeller is interested in the actual thermodynamic work being done by the fuel in such cases, or in the case of relatively weak driving, explicit modelling of the reverse reactions is unimportant. For strand displacement, with  a sufficiently long toehold, the equilibrium state is so biased towards the product of displacement that reverse reactions, leading to a residual concentration of the input, can be neglected.  In many other contexts, however, it is important to include microscopically reversible transitions, because the free energy of transitions is relatively weak (as is the case of self-assembly near the melting temperature); because the overall thermodynamics is of interest to the modeller; or because reverse reactions, despite being typically slow, have a profound impact on system behaviour. 

\begin{figure}
  \includegraphics[width=0.48\textwidth]{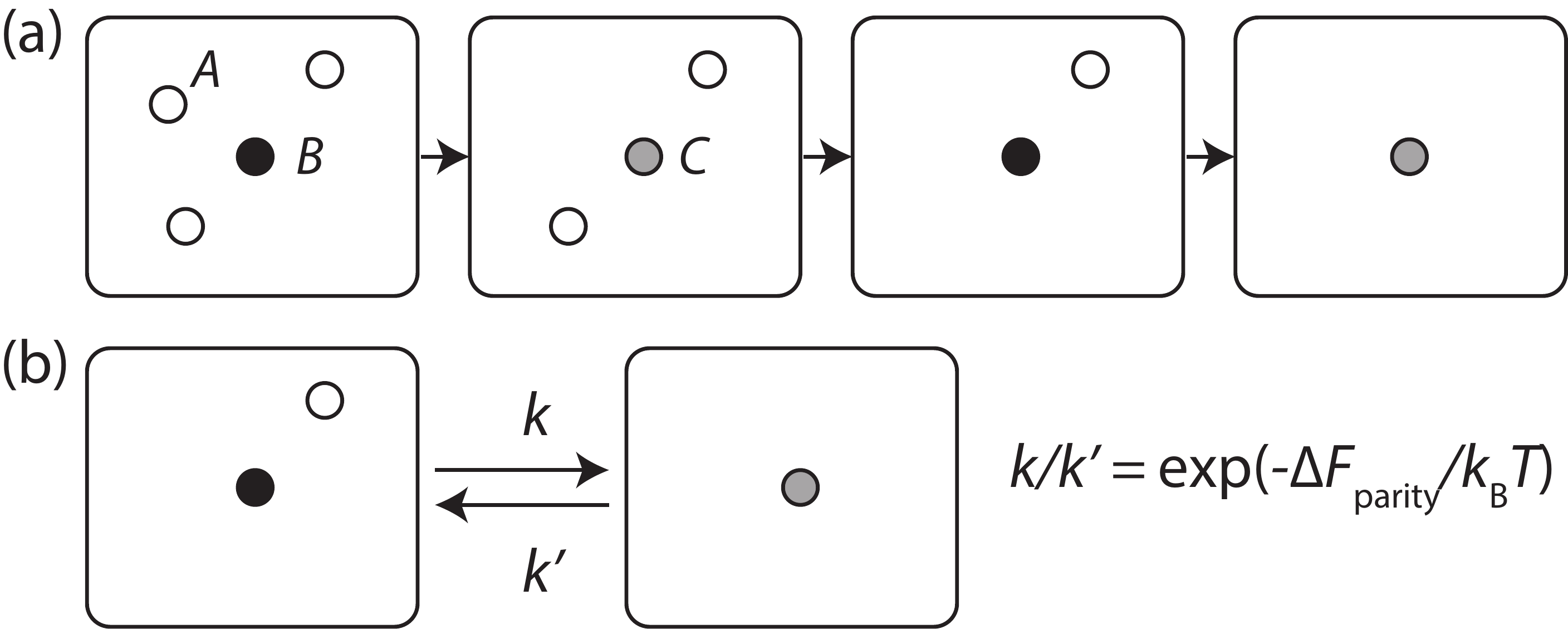}
\caption{Schematic illustration of the parity-computing algorithm of Cummings {\it et al.} \cite{Cummings2015}. (a) Reactions consume species $A$, and switch $B$ into $C$ and vice versa. Thus a system initated with an odd number $N_A^0$ of $A$ and a single $B$ will result in an isolated $C$ (as shown). Alternatively, a system initiated with an even number $N^0_A$ of $A$ molecules will result in a single isolated $B$. (b) The consequence of microscopic reversibility is a finite backwards rate for all transitions; we take the overall difference in free energy between $N_A=1$ and $N_A=0$ to be $\Delta F_{\rm parity}$}
\label{fig:parity}       
\end{figure}

For example, in the field of molecular computation and algorithms, it is common to assume that reactions can be made totally irreversible \cite{Zechner2016,Chen2013,Cummings2015,Chen2012,Briat2016}. An example is an algorithm for computing the parity (even/odd nature) of an initial number of molecules of type $A$, as discussed by Cummings {\it et al.} \cite{Cummings2015} and illustrated in Fig.~\ref{fig:parity}\,(a). The algorithm introduces two other species, $B$ and $C$, and the (assumed microscopically irreversible) reactions
\begin{align}
A + B \rightarrow C,  \nonumber \\
A + C \rightarrow B.
\end{align}
If the system is initially prepared with $N_A^0$ molecules of type $A$, and 1 molecule of type $B$, it can be seen that the reactions will interconvert $B$ and $C$ (retaining $N_B+N_C=1$) whilst reducing the number of $A$ molecules. The final state will be a single molecule of type $B$ if $N_A^0$ is even, and a single molecule of type $C$ if $N_A^0$ is odd. The output of the network is thus the state of the $B/C$ molecule in the limit of long time, which reports on the parity of $N_A^0$. Note that these reactions can in principle be implemented with DNA using a large  supply of implicit ancillary molecules \cite{Chen2013,Qian2011b}. 

What happens if we consider reverse reactions, so that $C \rightarrow A+B$ and $B \rightarrow A+C$ are also possible, as in Fig.~\ref{fig:parity}\,(b)? Then there is a finite chance that a system observed in the long-time limit will contain a $B/C$ molecule that does not reflect the parity of $N_A^0$. Specifically, let us assume that for a single $A$ and a single $B$ being converted into a single $C$ in the volume $V_\sigma$, in  the absence of all other molecules of type $A$, $B$ and $C$, the free-energy difference between macrostates is $F_\sigma(N_A=0, N_B=0, N_C=1) - F_\sigma(N_A=1, N_B=1, N_C=0) = \Delta F_{\rm parity}$ (Fig.~\ref{fig:parity}\,(b)). For simplicity, let us also assume that the same free-energy difference applies to  $A+C \rightarrow B$, $F_\sigma(N_A=0, N_B=1, N_C=0) - F_\sigma(N_A=1, N_B=0, N_C=1) = \Delta F_{\rm parity}$. These free energies include the contributions from any implicit ancillary molecules. 
Provided the ancillary molecules are in excess (their concentrations are essentially unaffected by the reactions involving $A$, $B$ and $C$), then the probability of observing $N_A$ molecules of type $A$ in equilibrium is given by
\begin{align}
&P^{\rm eq}_\sigma(N_A=1) =  P^{\rm eq}_\sigma(N_A=0) \exp(-\Delta   F_{\rm parity}/{k_{\rm B}}T), \nonumber\\
&P^{\rm eq}_\sigma(N_A=2) = \frac{1}{2} P^{\rm eq}_\sigma(N_A=0)\exp(-2\Delta   F_{\rm parity}/{k_{\rm B}}T),\nonumber\\
&P^{\rm eq}_\sigma(N_A=3) = \frac{1}{6} P^{\rm eq}_\sigma(N_A=0)\exp(-3\Delta   F_{\rm parity}/{k_{\rm B}}T), \nonumber\\
&P^{\rm eq}_\sigma(N_A) = \frac{1}{N_A!}P^{\rm eq}_\sigma(N_A=0) \exp(-N_A\Delta   F_{\rm parity}/{k_{\rm B}}T).
\label{eq:parity}
\end{align}
Each term includes an additional factor, ${\rm e}^{-\Delta   F_{\rm parity}/{k_{\rm B}}T}$, to account for the additional reaction that must take place to reach $N_A=0$. The $N_A!$ factor accounts for the fact that the free energy difference between states with $N_A$ and $N_A-1$ depends on the number of $A$ present, as we previously saw in Section~\ref{sec:free_energies}. One way to confirm the exact dependence is to note that the rate for $C \rightarrow A+B$ and $B \rightarrow A+C$  should be independent of $N_A$, but $A+B \rightarrow C$ and $A+C \rightarrow B$ should occur with an overall rate proportional to $N_A$. Incorporating this into the free energy gives the factors in Eq.~\ref{eq:parity} (a logarithmic growth in the free energy difference with $N_A$).

When $N_A$ is odd, the readout from the reporter molecule $B/C$ gives an incorrect readout for the parity of $N_A^0$. It can be seen that the even and odd terms of Eq.~\ref{eq:parity} correspond to terms in the expansion of hyperbolic functions of  ${\rm e}^{(-\Delta F_{\rm parity}/k_{\rm B}T)}$. Thus
\begin{equation}
\frac{P_{\rm correct}}{P_{\rm incorrect}} = \frac{P^{\rm eq}_\sigma(N_A\,\,{\rm even})}{P^{\rm eq}_\sigma(N_A\,\,{\rm odd})} \approx \coth \left({\rm e}^{(-\Delta F_{\rm parity}/k_{\rm B}T)}\right).
\end{equation}

For $\Delta F_{\rm parity}$ large and negative -- when the interconversion of an isolated $A$ and $B$ into an isolated $C$ is favourable -- the algorithm is accurate. For lower values of $\Delta F_{\rm parity}$, the accuracy is reduced. Of course, in an abstract design it is possible to imagine $\Delta F_{\rm parity}$ is as large as possible, but the need to do this should be noted. It is also worth highlighting the fact that  $\Delta F_{\rm parity}$ is {\it not} the standard free energy of the reaction; it is the free energy difference between a macrostate with a single $A$ and a single $B$ and a macrostate with a single $C$, in volume $V_\sigma$. In fact, it can be shown that \cite{Ouldridge_bulk_2010},
\begin{align}
\Delta F_{\rm parity} &= - k_{\rm B} T \ln \left(  \frac{K^{\rm eq}_{A+B}}{V_\sigma} \right) \nonumber \\
& =  k_{\rm B} T \ln \left(  \frac{V_\sigma}{V_0}  \right) +\Delta_{A+B} F^0,
\end{align}
where $\Delta_{A+B} F^0$ is the standard free energy of reaction, calculated in the reference volume $V_0 = 1/\mathcal{C}_0$.
Thus the larger the system volume $V_\sigma$, the larger the standard free energy must be to give the same accuracy. If $N_A^0$ is large, $V_\sigma$ will also have to be large to ensure that the system is dilute and functions as intended. Therefore the standard free energy required for accurate computation will be very large and negative, and will become more negative logarithmically with system size at fixed initial concentration $\mathcal{C}_A$. The robustness of other molecular algorithms to finite reverse reaction rates, and possible ways to mitigate these effects, remain important open questions -- although it seems likely that the least robust algorithms will be those that require a specific macrostate as an output.

\subsection{The meaning of thermodynamic reversibility}
\label{sec:thermo_reversibility}
\begin{figure}
  \includegraphics[width=0.48\textwidth]{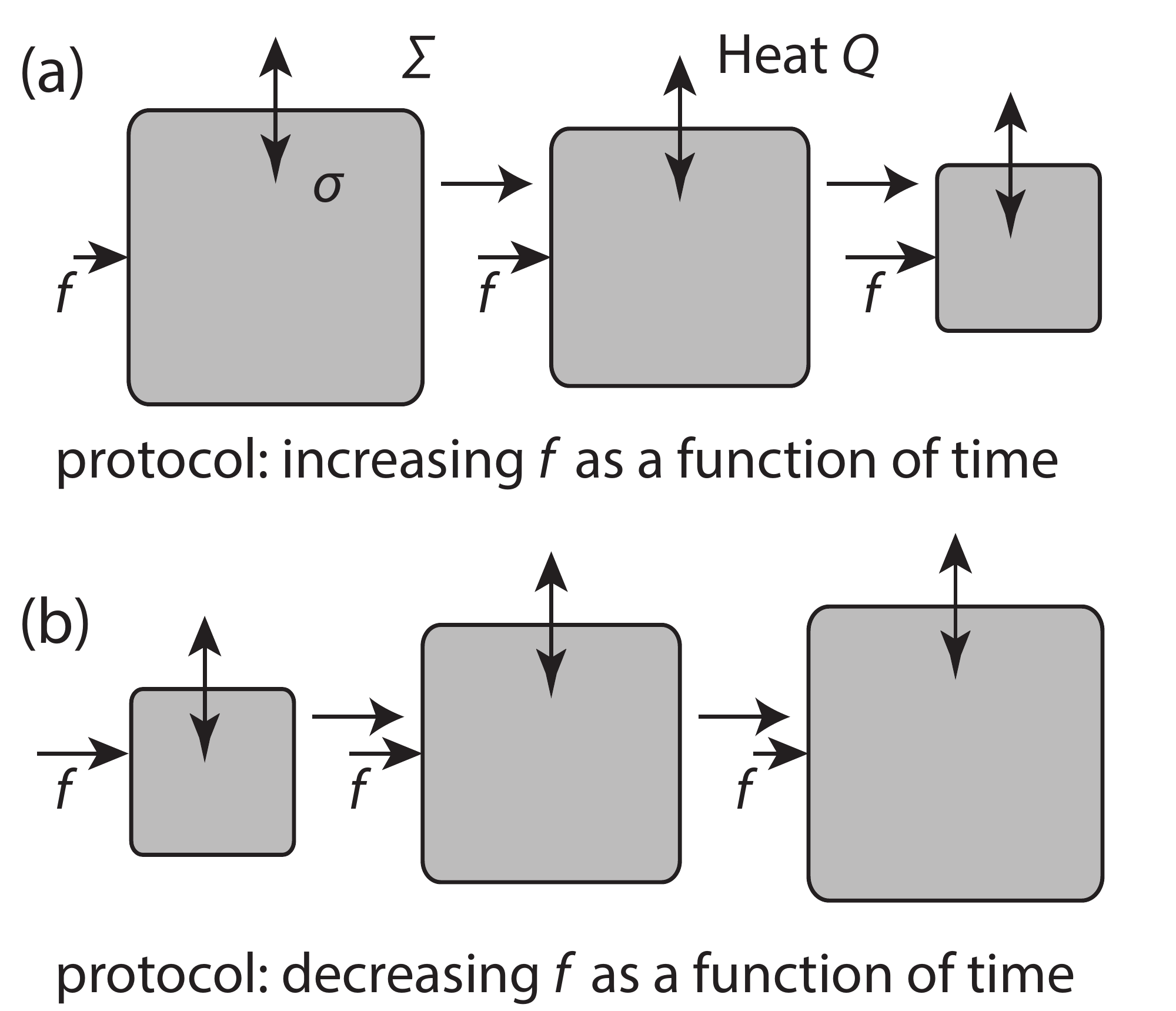}
\caption{The meaning of a reversible process. A schematic illustration of a process in which a time-dependent protocol (perhaps a compressing force) is applied to $\sigma$, leading to change in $\sigma$ and exchange of heat with the environment $\Sigma$. The process is reversible if applying a time-reversed protocol as in (b) leads to both $\sigma$ and $\Sigma$ being returned to their initial conditions.}
\label{fig:reversible}       
\end{figure}
 The microscopic reversibility of individual transitions should {\em not} be conflated with the idea of thermodynamic reversibility, despite the unfortunate similarity of nomenclature. Thermodynamic reversibility is not the property of an individual transition or even a set of rate parameters describing a system. Rather, it is a property of an entire process in which an experimentalist or machine in the environment manipulates the system from the outside, applying some protocol (changing the conditions with time, as shown in Fig.~\ref{fig:reversible}\,(a)).  If the system and environment are {\em both} be restored to their initial conditions by a time-reversed protocol (Fig.~\ref{fig:reversible}\,(b)), then the process and its time-reversed counterpart are thermodynamically reversible \cite{Adkins1987}. 

Reversible processes are important in thermodynamics because they do not increase the entropy of the universe, given by the sum of the entropy of $\sigma$ and its environment $\Sigma$  \cite{Adkins1987}. The entropy of the universe (an isolated system) cannot decrease, and excess entropy generation corresponds to wasted effort (or work). Hence  reversible processes are the most efficient way of manipulating a system between a given start and end point.
 Historically, there has been significant interest in the minimal entropic cost of certain  computational procedures, particularly the manipulation of a single bit, and the possibility of reversible computing \cite{Bennett1982,Landauer1961,Sagawa2009}.

It is often proposed that molecular systems might be candidates for the construction of reversible computers \cite{Mehta2012,Thachuk2012,Thachuk2013}, and computational architectures with microscopically reversible dynamics have been analysed \cite{Thachuk2012,Thachuk2013}. Importantly, however, {\em microscopic reversibility  does not imply thermodynamic reversibility.} In truth, all systems possess microscopic reversibility -- ignoring it is simply a modelling assumption, as discussed in Section \ref{sec:reversibility}. Thermodynamic reversibility, however, depends on the initial conditions of the system, and the way in which a protocol is applied -- not the intrinsic properties of a transition.

Consider, for example, a DNA-based toehold-exchange reaction (Fig.~\ref{fig:strand_exchange_markov}). We might be interested in switching the substrate strand $A$ from a $B$-bound state to a $C$-bound state. If we set up the system to ensure that we start with a single $AB$ duplex, and a separate $C$ strand, and let it evolve naturally ({\t ie., implement a trivial protocol of ``do nothing") the system will undergo repeated strand exchange reactions, and the final state will alternate between $AB$ and $AC$. This isn't a full switch from $AB$ to $AC$, but it is at least a change in the probability distribution over macrostates $P_\sigma(i)$. 

During this process, both forwards and backwards reactions occur, and continue to occur indefinitely. So is this a thermodynamically reversible process? It is not. If we were to reverse our protocol of ``doing nothing", the system would not return to a state in which it was guaranteed to have an $AB$ duplex; it would stay in an uncertain $AB$/$AC$ state. Indeed, any overall change of a system ({\it ie.,} a change in the probability distribution $P_\sigma(i)$) that occurs during a trivial ``do nothing" protocol is {\em necessarily thermodynamically irreversible}, since the opposite change in $P_\sigma(i)$ {will not} occur under a time-reversed ``do nothing" protocol. The actual entropy increase in the process can be calculated using the methods discussed in Section~\ref{sec:stoch_thermo}.


Note that it is not really meaningful to describe the individual reactions as reversible or irreversible in the thermodynamic sense. Once the system has relaxed to a completely uncertain equilibrium $AB$/$AC$ state,  strand exchange reactions will still occur from the perspective of individual trajectories. However, $P_\sigma(i)$ will not change, and the subsequent (trivial) evolution of the system is reversible. It is only the initial period of transitioning from a guaranteed $AB$ state to an uncertain $AB/AC$ state that is irreversible -- despite the fact that it involves the same microscopic strand exchange reactions. Fundamentally, thermodynamic reversibility is a property of the overall change of the system $\sigma$ (through $P_\sigma(i)$) and the environment $\Sigma$, rather than the individual molecular processes involved. 

\begin{figure}
  \includegraphics[width=0.48\textwidth]{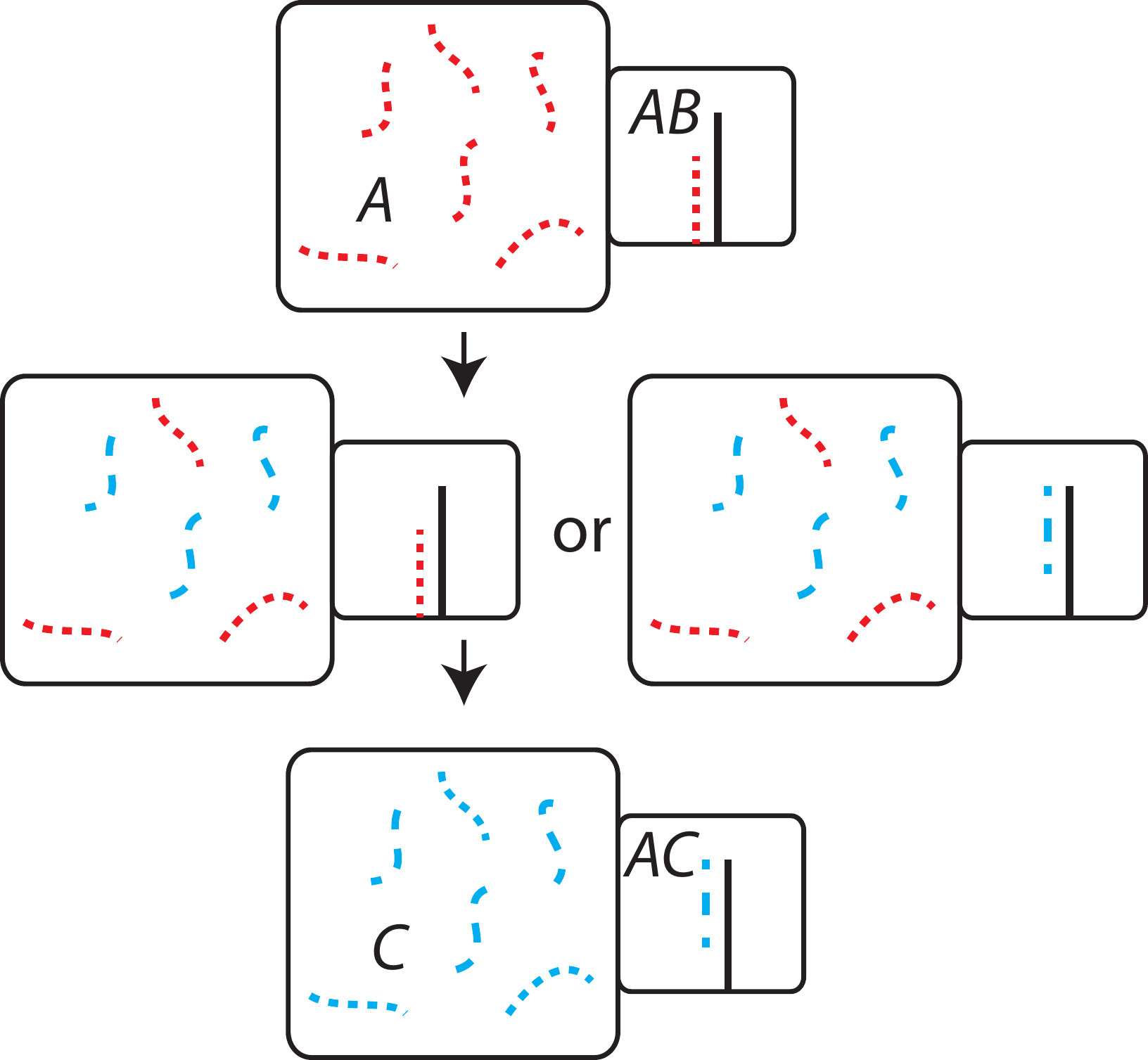}
\caption{A reversible switch of toehold-exchange system, achieved by coupling the volume containing strand $A$ to buffers containing increasing amounts of strand $C$, and decreasing amounts of strand $B$. If this process is performed gradually enough, it is reversible.}
\label{fig:reversible2}       
\end{figure}

Does this observation prohibit reversible operations  with molecular systems? In fact,  it is possible to drive a change in the state of molecular systems reversibly. However, this must be done by externally changing the solution conditions in a quasistatic (slow) manner. For the above example, a reversible switch could be achieved by initially coupling the $A$ strand to a buffer containing only $B$ strands, and then replacing this buffer with a series of alternatives with gradually increasing ratios of $\mathcal{C}_C/\mathcal{C}_B$, as shown in Fig.~\ref{fig:reversible2}. Eventually, in the limit $\mathcal{C}_C/\mathcal{C}_B \rightarrow \infty$, the switch to $AC$ will be complete. By reversing the protocol, we return to the initial $AB$ configuration, and restore the buffers to their original condition -- hence the process is reversible. Similar procedures for reversible copying of a bit and reversible construction of a polymer copy from a template are discussed in more detail in Ref. \cite{Ouldridge_copy_2015} and Ref. \cite{Ouldridge_copy_2016}, respectively. Although perhaps impractical, they highlight the difference between a thermodynamically reversible operation, and an irreversible process in which microscopic reversibility is relevant.

\section{Catalysis and the consumption of molecular fuel}
\label{sec:catalysis}
{Unlike the systems considered hitherto, living organisms do not rapidly tend towards equilibrium. Instead, they are kept out of equilibrium by a continuous supply of chemical fuel. Conceptually, we might imagine that this fuel is supplied from an enormous buffer that is not depleted on the time scale of interest, or perhaps that the fuel levels are continuously topped up by some process. Staying out of equilibrium is of course essential for living systems, since equilibrium systems are active. In this Section we will explore some basic functionalities of systems that are continuously supplied with fuel molecules. We will see that continuous fuel consumption allows systems to establish  non-equilibrium steady states. These  steady-states can, for example, store the electrochemical free energy needed to propagate nerve signals, and permit molecular signalling without consumption of the upstream signalling molecule.} 



\begin{figure}
  \includegraphics[width=0.48\textwidth]{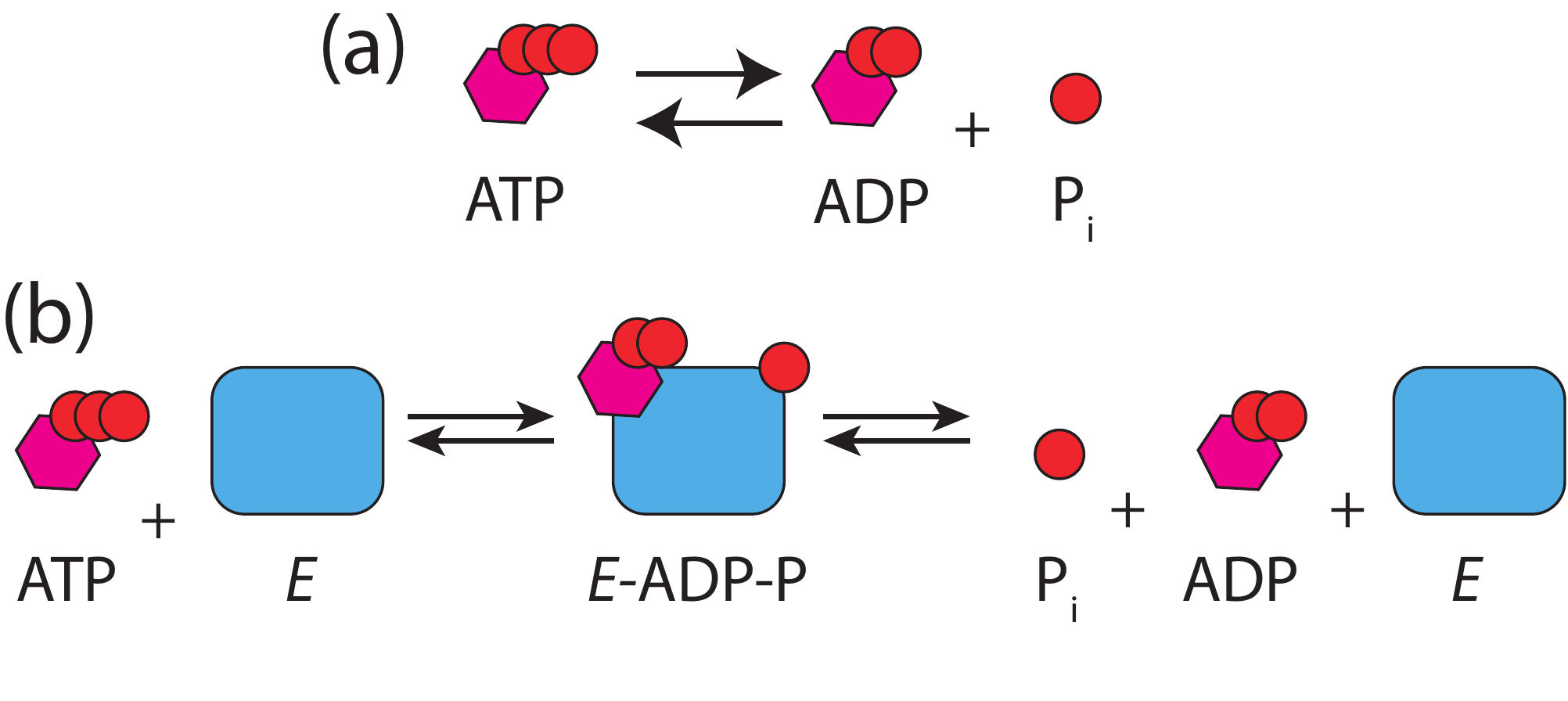}
\caption{(a) The breakdown of ATP into ADP and P$_{\rm i}$. (b) a catalyst $E$ can enhance the reaction rate by providing an alternative pathway; the catalyst is not consumed by the overall reaction.}
\label{fig:ATP}       
\end{figure}

The prototypical example of a chemical fuel molecule is ATP, and so we will  base our discussion around it. ATP consists of a sugar-base group attached to a chain of three phosphate groups \cite{Alberts2002}. All of these bonds are covalent, but phosphate groups can be removed by hydrolysis. In particular, a key reaction is the removal of phosphate to give ADP (Fig.~\ref{fig:ATP}\,(a)),
\begin{equation}
{\rm ATP} \rightleftharpoons {\rm ADP} +{\rm P}_{\rm i}.
\label{eq:ATP_hydrolysis}
\end{equation}
This reaction has an associated equilibrium constant $K^{\rm eq}_{\rm ADP+P_i}$ and a standard free energy of ATP formation $\Delta_{\rm ADP+P_i} F^0$. The cell uses the breakdown of food molecules to maintain a concentration imbalance of ATP, ADP and P$_{\rm i}$; {\it i.e.,}
\begin{equation}
\frac{\mathcal{C}_{\rm ATP}}{\mathcal{C}_{\rm ADP} \mathcal{C}_{\rm P_{i}}} > K^{\rm eq}_{\rm ADP + P_i},
\end{equation}
or equivalently 
\begin{align}
\Delta_{\rm ADP+P_i} F_\sigma &= \mu^{\rm ATP}_\sigma -(\mu^{\rm ADP}_\sigma + \mu^{\rm P_i}_\sigma) \nonumber \\
&= \Delta_{\rm ADP+P_i} F^0+ k_{\rm B}T \ln \left( \frac{\mathcal{C}_0 \mathcal{C}_{\rm ATP}}{\mathcal{C}_{\rm ADP} \mathcal{C}_{\rm P_i}} \right) \nonumber \\
& > 0.
\end{align}
Left in isolation, a solution of ATP, ADP and P$_{\rm i}$ prepared with these concentrations would relax to equilibrium by converting ATP into ADP and P$_{\rm i}$. This process, which requires the disruption of covalent bonds, has extremely slow kinetics. Consequently the cell is able to build up and maintain a large concentration imbalance. 

The kinetics of processes such as that in Eq.~\ref{eq:ATP_hydrolysis} can be accelerated by catalysts \cite{Nelson2004,Alberts2002}. Catalysts provide an alternative reaction pathway between the endpoints, {which might (for example) lower} the barriers associated with the formation and disruption of covalent bonds, {as in ATP hydrolysis}.  A schematic illustration of catalyst operation is given in Fig.~\ref{fig:ATP}\,(b), and one might record the effective reaction as
\begin{equation}
{\rm ATP} + E \rightleftharpoons E-{\rm ADP}-{\rm P_i} \rightleftharpoons {\rm ADP} +{\rm P}_{\rm i} +E
\end{equation}
with $E$ being the catalyst molecule. Importantly, the catalyst is released at the end of the process unchanged, and does not contribute to the free energy of the overall reaction (its stoichiometric coefficient is zero). Therefore the catalyst does not affect the equilibrium balance between its substrate molecules, it only accelerates the rate at which this equilibrium is reached \cite{Nelson2004}. An equivalent way to view the action of a catalyst is that it accelerates {\em both} forwards and backwards rates equally. 

In nature, some catalysts are simply present to accelerate reaction kinetics. For example, amylase is present to accelerate the breakdown of starch into sugars. In principle, it can also accelerate the conversion of sugars into starch, but this is irrelevant in the high starch conditions in which it operates. Many catalysts act on the ATP hydrolysis reaction in Eq.~\ref{eq:ATP_hydrolysis}, but simply accelerating the equilibration of this process serves no function. Doing so would merely reduce the concentration imbalance built up by the cell. The purpose of these catalysts is to couple ATP hydrolysis to other reactions through the details of their internal biochemistry, and thereby use the free energy of ATP hydrolysis to drive those reactions in one way or another.

\begin{figure}
  \includegraphics[width=0.48\textwidth]{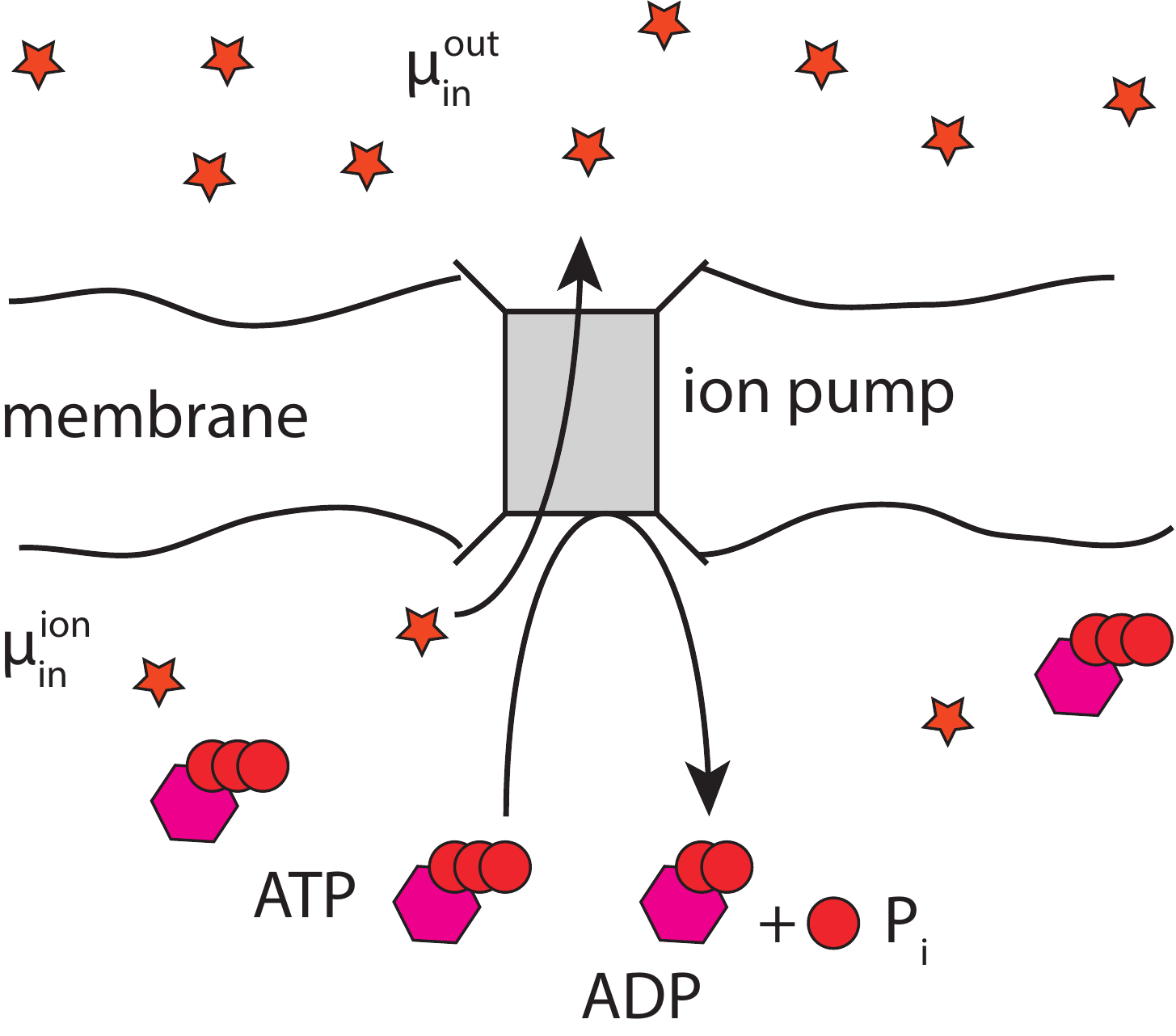}
\caption{An illustration of the principle of an ion pump. The pump systematically drives ions from inside the membrane to outside. By coupling the ion transition to ${\rm ATP} \rightleftharpoons {\rm ADP} +{\rm P}_{\rm i}$, ions can be systematically pumped against a chemical potential bias $\mu^{\rm ion}_{\rm out} - \mu^{\rm ion}_{\rm in}\geq0$. This pumping is possible due the large and negative contribution of the breakdown of ATP to the free energy difference associated with the transition.}
\label{fig:ion-pump}       
\end{figure}

A clear example is given by ion pumps, molecular machines that reside in membranes, and transport ions across these membranes (Fig.~\ref{fig:ion-pump}). Without coupling to ATP hydrolysis, these ion pumps could only facilitate the equilibration of ions on either side of the membrane ({\it i.e.}, allow current to flow until the chemical potential of the ions is equal on both sides; $\mu^{\rm ion}_{\rm in} = \mu^{\rm ion}_{\rm out}$). If, however, outwards ion transport is tightly coupled to ATP breakdown, then the overall free energy of pumping an ion outward is
\begin{equation}
\Delta_{\rm in\,+\,ATP} F_\sigma = \mu^{\rm ion}_{\rm out} - \mu^{\rm ion}_{\rm in} - (\mu_\sigma^{\rm ATP} -(\mu_\sigma^{\rm ADP} + \mu_\sigma^{\rm P_i})),
\end{equation}
which can be negative even if $\mu^{\rm ion}_{\rm in} - \mu^{\rm ion}_{\rm out}  > 0$, due to the imbalance of chemical fuel molecules maintained by the cell.
If the non-equilibrium chemical potentials of the ATP, ADP and P$_{\rm i}$ are maintained indefinitely, the ions will eventually reach a non-equilibrium steady state in which the outwards pumping is balanced by leaks back through other channels.

The non-equilibrium initial state of the fuel molecules is effectively a store of useful work. This store can be consumed to drive another process -- in this case, the  transfer ions against their chemical potential bias. The maximum chemical potential bias $\mu^{\rm ion}_{\rm in} - \mu^{\rm ion}_{\rm out}$ against which progress can be made  is simply the excess free energy change associated with breakdown of a single ATP, $\mu_\sigma^{\rm ATP} -(\mu_\sigma^{\rm ADP} + \mu_\sigma^{\rm P_i})$. 

Transferring ions across membranes is very much like charging a capacitor, and cells use these capacitors to drive other processes, including the firing of nerve cells. There are numerous other natural processes in which molecular fuel consumption is used to drive a coupled reaction. Molecular motors such as myosin, kinesin and dynein catalyse ATP breakdown to bias the direction in which they walk along a track \cite{Nelson2004,Howard2001,Alberts2002}. Without coupling the forwards step to ATP hydrolysis, forwards and backwards steps would be equally likely, since all walker binding sites on the track have equal free energy by symmetry. 

By coupling continuous fuel consumption to other molecular reactions via catalysts, it is therefore possible to drive those other reactions away from their natural equilibrium \cite{Qian2007,Beard2008}. A particularly important  example are push-pull motifs, {which} are ubiquitous in the signalling mechanisms which pass information around the cell and illustrated in Fig.~\ref{fig:pushpull}. In these 
small networks, the presence of active upstream catalysts leads to activation of downstream substrates, which can  in turn propagate or respond to the signal.

Each push-pull motif consists of a protein that can be switched between its active ($X^*$) and inactive ($X$) states by binding of P$_{\rm i}$ to one (or more) amino acid residues. If this ``phosphorylation" could only occur through binding and unbinding from P$_{\rm i}$ in solution, then the activity level would swiftly tend towards that determined by the equilibrium constant of binding and the abundance of  P$_{\rm i}$,
\begin{equation}
\frac{\mathcal{C}^{\rm eq}_{X^*}}{\mathcal{C}^{\rm eq}_X} = \mathcal{C}_{\rm P_{i}} K^{\rm eq}_{X+{\rm P_i}}
\end{equation}
Importantly, any upstream catalyst  could only enhance this convergence to equilibrium, it could not change it. As discussed in Section~\ref{sec:self-assembly}, relative equilibrium abundances are determined exclusively by the free energies of reactants and products -- the concentration of a transiently involved catalysts is irrelevant. 

\begin{figure}
  \includegraphics[width=0.48\textwidth]{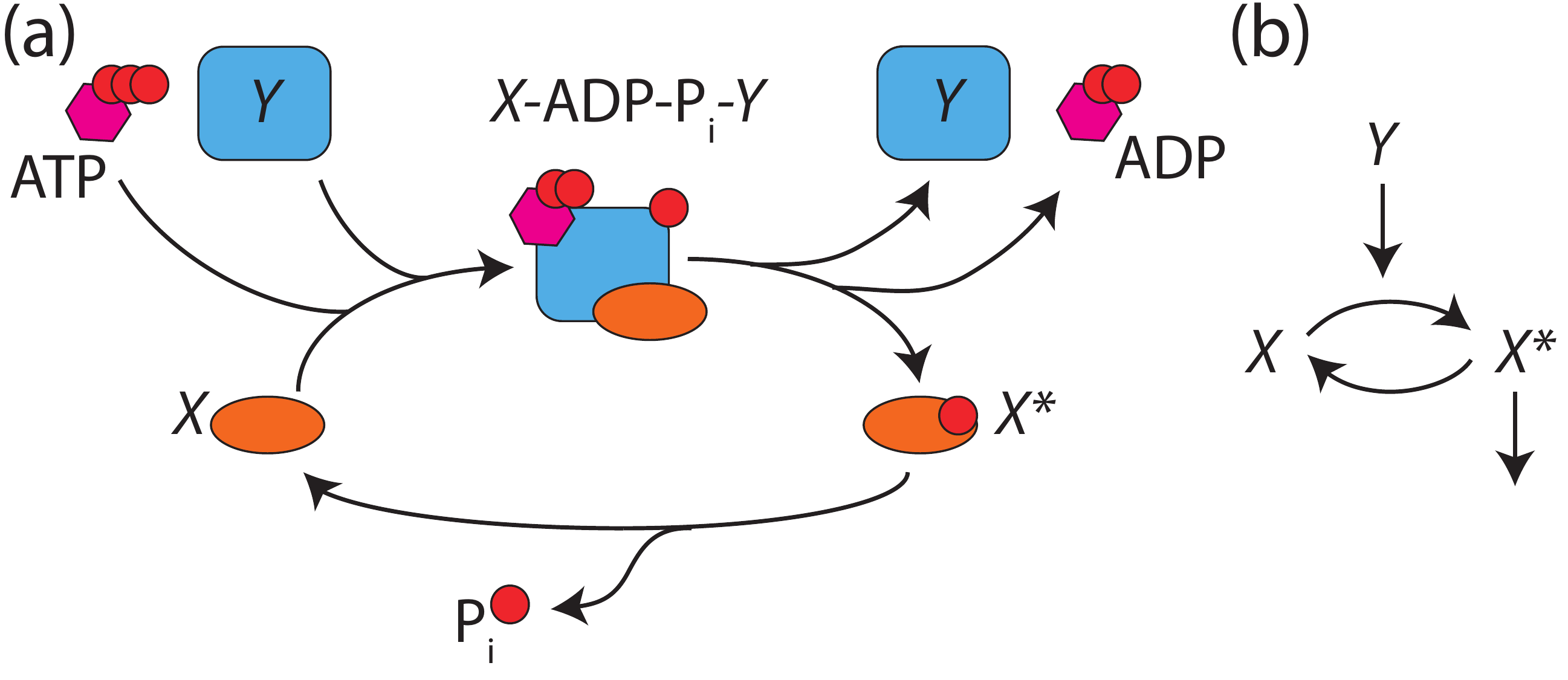}
\caption{A push-pull motif. The downstream molecule $X$ can be converted between its phosphorylation states by two pathways; by exchanging phosphate with ADP/ATP or solution. Catalysts can accelerate one or both of these reactions; in (a), a catalyst $Y$ is shown accelerating interactions with ATP/ADP.  The overall negative free energy change of ATP breakdown favours the direction of reactions shown, rather than their reversed counterparts, and so $X/X^*$ effectively undergoes a cycle. As a result, a signal related to the concentration of $Y$ can be passed onto $X^*$, which then propagates the signal further. This function of the push-pull motif is illustrated in the simplified diagram in (b). }
\label{fig:pushpull}       
\end{figure}

However, if ATP is  present, phosphorylation can also occur via transfer of phosphate from ATP:
\begin{equation}
X + {\rm ATP} \rightleftharpoons X^* + {\rm ADP}.
\end{equation}
If the ATP, ADP and  P$_{\rm i}$ concentrations are maintained such that $\Delta_{\rm ADP+P_i} F_\sigma >0$, it is impossible for the $X/X^*$ subsystem to reach equilibrium. The series of reactions shown in Fig.~\ref{fig:pushpull}\,(a)
\begin{equation}
X + {\rm ATP} \rightarrow X^* +{\rm ADP} \rightarrow X + {\rm ADP} + {\rm P_i},
\label{eq:forward_cycle}
\end{equation}
will necessarily occur much more frequently on sample trajectories than the counterpart
\begin{equation}
 X + {\rm ADP} + {\rm P_i} \rightarrow X^* +{\rm ADP} \rightarrow  X + {\rm ATP}.
\label{eq:backward_cycle}
\end{equation}
The reactions in Eq.~\ref{eq:forward_cycle} will occur more frequently since the net result of the reactions in Eq.~\ref{eq:forward_cycle} is the breakdown of ATP, whereas the net result of the reaction in Eq.~\ref{eq:backward_cycle} is the synthesis of ATP from ADP and P$_{\rm i}$, and the sign of $\Delta_{\rm ADP+P_i} F_\sigma = \mu_\sigma^{\rm ATP} -(\mu_\sigma^{\rm ADP} + \mu_\sigma^{\rm P_i}) >0$ favours ATP breakdown. The protein  $X$ systematically tends to be converted into $X^*$ by one pathway, and converted back into $X$ by a completely different pathway (Fig.~\ref{fig:pushpull}\,(a)). From the perspective of $X/X^*$, detailed balance appears to be violated, and therefore the coupling to chemical fuel drives the $X/X^*$ subsystem out of equilibrium.  As a result, when the hydrolysis-driven cycle of Eq.~\ref{eq:forward_cycle} becomes dominant relative to Eq.~\ref{eq:backward_cycle}, the steady state ratio $\mathcal{C}_{X^*}/\mathcal{C}_{X}$ is determined purely by the relative rates of the two reactions within that cycle, $X + {\rm ATP} \rightarrow X^* +{\rm ADP} $ and $  X^* +{\rm ADP}  \rightarrow X+ {\rm ADP} + {\rm P_i}$.

Since the dominant reactions that interconvert $X$ and $X^*$ are not a microscopic reverse pair, thermodynamics places no restrictions on their relative rates. In particular, it is perfectly possible for a catalyst to accelerate the kinetics of one reaction and not the other. The output ratio $\mathcal{C}_{X^*}/\mathcal{C}_{X}$ can then be sensitive to the concentration  of an upstream catalyst ($Y$ in Fig.~\ref{fig:pushpull}\,(a)) that accelerates the exchange of phosphate between $X$ and ATP. In this manner, signalling cascades can pass on information on the concentration of active upstream catalysts, illustrated schematically in Fig.~\ref{fig:pushpull}\,(b), as required. We emphasize that it would be impossible for $Y$ to influence the $X/X^*$ ratio in this way without consuming fuel, or binding to the downstream molecule and remaining bound ({\it ie., with $Y$ itself being consumed by the reaction}). 

{We emphasize that the  violation of detailed balance for $X$/$X^*$ only occurs because the fuel molecules are continuously being consumed, which prevents the $X/X^*$  system from reaching equilibrium. If the root source of the chemical fuel was also explicitly modelled as part of an extended description, we would see the system  as a whole relaxing towards equilibrium through the reactions that turn over ATP;   the apparent non-equilibrium steady state of $X/X^*$ is just a psuedo-steady state that arises because the dynamics of $X/X^*$ are much faster than that of the extended system as a whole. Furthermore, the fundamental relationship between transition rates derived from detailed balance, Eq.~\ref{eq:db}, still holds: the relative rate of each microscopic reverse pair of forward and backward transitions is still determined by the underlying $\Delta F$ in the usual way, with the contributions from the fuel molecules to $\Delta F$ taken into account.}

The above catalytic activation motif is extremely powerful. Since the upstream molecule acts catalytically, it is not consumed in the act of passing on the signal. It is thus able to interact with other downstream proteins to either amplify or branch the signal \cite{Mehta2016}. Moreover, the downstream readout's persistent modification allows it to keep a long-lived record of the state of the upstream protein, permitting time-integration of signals \cite{Govern_PRL_2014,govern2014}. These features are only
possible in a motif that incorporates catalytic signal propagation, which in turn relies on coupling to a non-equilibrium fuel source.

Clearly, driving of reactions by coupling them to a non-equilibrium fuel supply is a central motif in natural systems \cite{Nelson2004,Qian2007,Beard2008}; other examples include enhancement of substrate selectivity through ``kinetic proofreading"   \cite{Hopfield1974,Ninio1975}, important in replication, transcription and translation of nucleic acids; and maintenance of oscillations related to cell cycle \cite{Tostevin2006} and circadian clocks \cite{Paijmans2016}. Needless to say, these fuel-consuming processes are widely exploited in synthetic biology, and enzyme-driven processes in cell-free environments \cite{Fujii2012,Green2014,Stricker2008}.  There has also been some exploration of these ideas in nucleic acid  nanotechnology, most notably in the design of autonomous DNA walkers \cite{Bath2005,Bath2009,Wickham2011,Tian2005,Lund2010,Cha2014,Muscat2011}, some of which are powered by base pairing alone  \cite{Muscat2011,Green2008,Omabegho2009}. In particular, hairpins can be used as metastable non-equilibrium fuel \cite{Turberfield2003,Green2008,Omabegho2009}; the motion of the walker is coupled to catalysing reactions that are otherwise frustrated by the hairpin structure.

 More generally, recent work has shown how to implement arbitrary chemical reaction networks (CRNs) as nucleic acid systems by realising each reaction as a multi-stage process involving ancillary molecules \cite{Chen2013,Qian2011b}. These ancillary molecules can function as fuel, in principle allowing catalytic driving out of equilibrium as discussed above \cite{Chen2013}. Most impressively, the Khammash group using a nucleic-acid-based architecture to implement a CRN-based noise filter  powered by fuel consumption \cite{Zechner2016}. When considering the capabilities and design possibilities of CRNs, it is always worth noting whether non-equilibrium fuel is required to provide a large thermodynamic bias for the system to function as intended. Furthermore, when these implicit fuel molecules are systematically consumed by the system, detailed balance will in general be violated for the other species considered explicitly. 

Notwithstanding the above examples, the common natural motif of a network powered by fuel consumption is relatively rare in non-enzymatic artificial nanotechnology. Given the uses of fuel-powered systems, producing artificial analogues is an obvious goal. One major advantage would be that, given a constant supply (or sufficiently large buffer) of fuel, such systems could potentially operate indefinitely, performing repeated operations, rather than functioning as single-shot devices. Continuous operation would be essential for implementing circuits that perform functions like feedback-control \cite{Briat2016}.

\section{Stochastic thermodynamics and its relevance to biochemical systems}
\label{sec:stoch_thermo}
{
Most interesting molecular systems  are out equilibrium. However, thus far we have only discussed thermodynamic quantities $U^{\rm eq}_\sigma$, $S^{\rm eq}_\sigma$ and $F^{\rm eq}_\sigma$ defined with respect to the equilibrium distribution. Macrostate free energies, energies and entropies, whilst useful, do not quantify the thermodynamic properties of a general non-equilibrium distribution. In this Section, we discuss how these thermodynamic ideas can be extended to individual stochastic trajectories and evolving non-equilibrium distributions over macrostates. In doing so, we will draw heavily upon the pioneering work of Crooks, Seifert, Esposito and Van den Broeck  \cite{Crooks1999,Esposito2011,Seifert2005}. 
}

{
Section~\ref{sec:thermo_reversibility} suggests that entropy generation is related to the relative probability of forward and reverse trajectories. To proceed with this line of reasoning, we let  $z(t)$ be a sample trajectory generated by the underlying Markovian dynamics in the space of macrostates of the system, with  initial value $z(0)$ an  final value $z(\tau)$. We then define $P_\sigma([z(t)]|z(0))$ as the probability of observing that trajectory given the initial condition, 
 and ${P}_\sigma([\tilde{z}(t)]|z(\tau))$ as the probability of observing exactly the time-reversed trajectory  given a starting point of $z(\tau)$. Note that if a time-dependent external protocol is applied as in Section~\ref{sec:thermo_reversibility}, that protocol must be time-reversed to calculate ${P}_\sigma([\tilde{z}(t)]|z(\tau))$.  We ignore such a possibility for our simple biochemical systems. 

The following sections follow from a single assumption quantifying the relationship between the relative likelihoods of ${z}(t)$ and its microscopic reverse $\tilde{z}(t)$, and entropy changes \cite{Crooks1999,Seifert2005,Jarzynski2011,Seifert2012}:
\begin{align}
k_{\rm B} \ln \frac{P_\sigma([z(t)]|z(0))}{{P}_\sigma([\tilde{z}(t)]|z(\tau))} &= \Delta S_{\Sigma}[z(t)] + (S_\sigma(z(\tau)) - S_\sigma(z(0)). \nonumber \\
\label{eq:db2}
\end{align}
Here, $\Delta S_{\Sigma}[z(t)]$ is the change in the entropy of the environment due to the trajectory $z(t)$, given by the heat deposited therein. $\Delta S_{\Sigma}[z(t)] = -\Delta U_\sigma[z(t)] /T$ in the absence of a time-dependent protocol, as we assume \cite{Seifert2011} (for these purposes, it is simplest to treat any fuel as  being supplied by a large buffer that is explicitly modelled part of the system  $\sigma$). Eq.~\ref{eq:db2}  holds very generally \cite{Crooks1999,Seifert2005,Jarzynski2011,Seifert2012}, and in particular is necessarily true for the simple molecular systems that we have considered. In fact, Eq.~\ref{eq:db2} follows fairly straight-forwardly from applying the fundamental relation derived from detailed balance, Eq.~\ref{eq:db}, at each sub-step of a trajectory.} 



\subsection{Generalising thermodynamic quantities to non-equilibrium distributions and fluctuating trajectories}
\label{sec:generalised_FE}
{
We are now ready to generalise the equilibrium quantities $U^{\rm eq}_\sigma$, $S^{\rm eq}_\sigma$ and $F^{\rm eq}_\sigma$ to non-equilibrium distributions over macrostates, and use Eq.~\ref{eq:db2} to derive constraints that show the power of these generalised quantities.}

The obvious generalisation of the internal energy to an arbitrary distribution $P_\sigma({\bf x},{\bf p})$ is
\begin{equation}
\mathcal{U}_\sigma[P_\sigma({\bf x},{\bf p})] =  \int_{{\bf x},{\bf p}} {\mathrm d}{\bf x},{\mathrm d}{\bf p}\,  E_\sigma({\bf x},{\bf p}) P_\sigma({\bf x},{\bf p}),
\end{equation}
the average of the microstate energy over $P_\sigma({\bf x},{\bf p})$. If the Markovian approximation for macrostate dynamics holds, then by definition the system is well-equilibrated within macrostates, even if different macrostates have non-equilibrium probabilities $P_\sigma(i) \neq P^{\rm eq}_\sigma(i)$. In this case,
\begin{equation}
\mathcal{U}_\sigma[P_\sigma(i)] =  \sum_i P_\sigma(i) U_\sigma(i).
\end{equation}
We also generalise the entropy in the same way, since the statistical quantity is well-defined for any distribution.
\begin{equation}
\mathcal{S}_\sigma[P_\sigma({\bf x},{\bf p})]= - k_{\rm B}\int_{{\bf x},{\bf p}} {\mathrm d}{\bf x},{\mathrm d}{\bf p}\, P_\sigma({\bf x},{\bf p}) \ln ( P_\sigma({\bf x},{\bf p})/\rho).
\end{equation}
When considering well-defined macrostates, this expression becomes
\begin{equation}
\mathcal{S}_\sigma[P_\sigma(i)] =  -k_{\rm B}\sum_i P_\sigma(i)\ln P_\sigma(i) + \sum_i P_\sigma(i) S_\sigma(i).
\label{eq:generalised_S}
\end{equation}
The generalisation is slightly more complex than for the internal energy; we obtain a term corresponding to the average entropy of the macrostates {\em and} a term corresponding to the uncertainty in the macrostate. Similarly, we can generalise the free energy \cite{Esposito2011,Parrondo2015}:
\begin{align}
\mathcal{F}_\sigma[P_\sigma(i)] &= \mathcal{U}_\sigma[P_\sigma(i)] - T\mathcal{S}_\sigma[P_\sigma(i)] \nonumber \\
 &=   \sum_i P_\sigma(i) F_\sigma(i)  +  k_{\rm B} T \sum_i P_\sigma(i) \ln P_\sigma(i).
\label{eq:generalised_FE}
\end{align}
{
Finally, it is useful to define a ``trajectory-dependent" entropy production \cite{Seifert2005,Jarzynski2011,Seifert2012}:
\begin{align}
&\Delta {s}[z(t)] = \nonumber \\
&\Delta S_\Sigma[z(t)]   +  (S_\sigma(z(\tau)) - S_\sigma(z(0))  - k_{\rm B}\ln \frac{P_\sigma(z(\tau))}{P_\sigma(z(0))}.
\label{eq:traj_ent}
\end{align}
This exotic quantity, when averaged over all possible trajectories $z(t)$, will give the generalised entropy change of the entire process -- hence its name. To see that $\sum_{z(t)} P_\sigma[z(t)] \Delta {s}[z(t)] = \Delta \mathcal{S}_{\sigma + \Sigma}$, compare Eq.~\ref{eq:traj_ent} to Eq.~\ref{eq:generalised_S}. Note that Eq. ~\ref{eq:traj_ent} contains a term for the entropy of the environment $\Sigma$, a term for the change in macrostate entropy $S_\sigma$, and a term related to the distribution of $\sigma$ over its macrostates, as required. 
}

{
These generalised quantities are particularly useful due to the constraints placed on their evolution by Eq.~\ref{eq:db2}. In particular, 
arguably the deepest result of stochastic thermodynamics is ``deriving" the second law of thermodynamics for generalised entropies using Eq.~\ref{eq:db2}. Moreover, the second law  inequality follows from a {\it fluctuation relation} equality for the trajectory-dependent entropy, highlighting the underlying physics.  We will now briefly outline this derivation for the relevant case of a simple chemical system that can only exchange heat with its environment. The general case is more complicated, but the ideas are similar \cite{Crooks1999,Seifert2005,Jarzynski2011,Seifert2012}. 

Combining Eq.~\ref{eq:db2} and Eq.~\ref{eq:traj_ent}, we see that}
\begin{equation}
\frac{P_\sigma[z(t)]}{{P_\sigma}[\tilde{z}(t)]} \exp(-\Delta s[z(t)]/k_{\rm B}) =1.
\end{equation}
Here, $P_\sigma[z(t)]$ is the probability of observing the trajectory $z(t)$, including the initial probability of being at $z(0)$, and $P_\sigma [\tilde{z}(t)]$ is the probability of observing the reverse trajectory, including a distribution of initial states of the reverse trajectories given by $P_\sigma(z(\tau))$. We can multiply by ${P}_\sigma[\tilde{z}(t)]$ and sum over all possible trajectories, yielding
\begin{equation}
\sum_{z(t)} P_\sigma[z(t)] \exp(-\Delta s[z(t)]/k_{\rm B}) = \sum_{\tilde{z}(t)} P_\sigma[\tilde{z}(t)] =1,
\end{equation}
in which we have used the fact that a sum over all trajectories $z(t)$ is equivalent to a sum over all reverse trajectories $\tilde{z}(t)$.
We have arrived at the celebrated fluctuation relation for entropy \cite{Seifert2005,Jarzynski2011,Seifert2012},
\begin{equation}
\langle \exp(-\Delta s[z(t)]/k_{\rm B}) \rangle = 1,
\label{eq:FR}
\end{equation} 
where the average is defined over all possible trajectories $z(t)$.  The conventional second law then follows using Jensen's inequality
$
\ln \langle \exp(f(v)) \rangle \leq \langle f(v) \rangle
$,
which implies
\begin{align}
\langle \Delta s[z(t)]/k_{\rm B} \rangle &= \Delta \mathcal{S}_{\Sigma}[z(t)] +\mathcal{S}_\sigma[P_\sigma(\tau)] - \mathcal{S}_\sigma[P(0)] \nonumber \\
& = \Delta \mathcal{S}_{\sigma + \Sigma}[z(t)] \geq 0 
\end{align}
{or equivalently that the total entropy of system (calculated sing the generalised entropy) and environment cannot decrease  \cite{Seifert2005}. 
}

The fluctuation relation is a remarkable result, implying that at the level of individual trajectories, ``negative entropy" paths are possible but statistically unfavoured \cite{Seifert2005}. The result effectively shows that the familiar second law follows from a more general statement about the statistical properties of trajectories, which in turn follows directly from the relatively simple assumption related to microscopic reversibility (Eq.~\ref{eq:db2}). As a result, this fluctuation theorem emphasises the statistical interpretation of entropy and the second law, and their natural emergence from familiar stochastic dynamics. A host of alternative fluctuations can also be derived, depending on the context \cite{Crooks1999,Jarzynski2011,Seifert2012}.

{Given the definition in Eq.~\ref{eq:generalised_FE}, the generalised free energy of a simple chemical system plays the same central role as the chemical free energy in equilibrium systems \cite{Esposito2011,Parrondo2015}. Again using the fact that a simple chemical system can only exchange energy in the form of heat with its environment, $\Delta \mathcal{S}_{\rm \Sigma} =  -(\mathcal{U}[P^\prime(i)]- \mathcal{U}[P(i)])/T$ \cite{Seifert2011}. Thus
\begin{align}
T\Delta \mathcal{S}_{\sigma + \Sigma} &= T\Delta \mathcal{S}_{\Sigma} + T\mathcal{S}_\sigma[P_\sigma^\prime(i)]- T\mathcal{S}_\sigma[P_\sigma(i)] \nonumber \\
& = -(\mathcal{U}_\sigma[P_\sigma^\prime(i)]- \mathcal{U}_\sigma[P_\sigma(i)]) \nonumber \\
& + T\mathcal{S}_\sigma[P_\sigma^\prime(i)]- T\mathcal{S}_\sigma[P_\sigma(i)]  \nonumber\\
& = -(\mathcal{F}_\sigma[P_\sigma^\prime(i)]- \mathcal{F}_\sigma[P_\sigma(i)]).
\label{eq:total_ent}
\end{align}
 In more complex environments, alternative results are obtained which also included the external work done on the system  \cite{Esposito2011,Parrondo2015,Seifert2011}.
}

Since the total generalised entropy of Eq.~\ref{eq:total_ent} is guaranteed to be an non-decreasing function of time, we can rephrase the  second law for s non-equilibrium molecular processes (in the absence of external work) in its most useful form:
\begin{equation}
T\Delta \mathcal{S}_{\sigma+\Sigma}  = -(\mathcal{F}_\sigma[P_\sigma^\prime(i)]- \mathcal{F}_\sigma[P_\sigma(i)]) \geq 0.
\label{eq:generalised_entropy_change}
\end{equation}
Eq.~\ref{eq:generalised_entropy_change} is highly significant. Firstly, it emphasizes the importance of the (generalised) free energy as a thermodynamic resource in simple molecular systems. Since total $\mathcal{F}_\sigma[P_\sigma(i)]$ can only decrease, if a process acts to increase $\mathcal{F}_\sigma[P_\sigma(i)]$ for a subset of species, a compensatory decrease in $\mathcal{F}_\sigma[P_\sigma(i)]$ must occur elsewhere. This observation generalises the discussion in Section~\ref{sec:catalysis} on the use the high free energy of chemical fuel molecules to drive other components of a system out equilibrium.

{
\subsection{Applications of stochastic thermodynamics}
}
Many of the results of stochastic thermodynamics, including the fluctuation relation itself, are arguably more philosophically deep than immediately useful. However, the fluctuating, far-from-equilibrium nature of many biochemical systems -- particularly at the single molecule level -- often lends itself to analysis using these tools. {In this Subsection we discuss some characteristic examples.}

Firstly, the generalised free energy and entropy allow us to meaningfully analyse the entropy generation (and hence irreversibility) of processes such as that considered in Section~\ref{sec:reversibility}, when a duplex  initially prepared in the $AB$ state relaxes to an equilibrium of equal probability $AB/AC$, via repeated rounds of strand exchange. If the equilibrium distribution has $AB$ and $AC$ duplexes with equal probability (we assume that the strands are dilute enough that three-stranded complex is rarely observed), then both macrostates have equal free energy.  Thus $\sum_i P_\sigma(i) F_\sigma(i)$ is equal in the initial and final states. Any change in generalised free energy can only arise from the second term in Eq.~\ref{eq:generalised_FE}, the difference in generalised entropy due to the distribution over macrostates. Indeed, for a system initially guaranteed to be in macrostate $AB$,  $\sum_i P_\sigma(i) \ln P_\sigma(i) =0$; whereas the uncertainty in the final macrostate gives $-\sum_i P_\sigma^\prime(i) \ln P_\sigma^\prime(i) = \ln 2$. Consequently 
\begin{equation}
T\Delta \mathcal{S}_{\sigma + \Sigma}  = -(\mathcal{F}_\sigma[P_\sigma^\prime(i)]- \mathcal{F}_\sigma[P_\sigma(i)]) = k_{\rm B}T \ln 2 >0
\end{equation}
for the irreversible  strand exchange process discussed in Section~\ref{sec:reversibility}. Note that the full definition of the generalised free energy, including the term arising from the entropy of the distribution over macrostates, is necessary to obtain this result. 

Perhaps even more interesting is the case when neither the initial {\em nor} the final distribution correspond to equilibrium. For example, a self-assembling structure need not reach the equilibrium distribution after a finite time, implying  $ \mathcal{F}_\sigma[P^\prime_\sigma(i)]> \mathcal{F}_\sigma[P_\sigma^{\rm eq}(i)]$, since the equilibrium state  minimises the generalised free energy by definition. But since $ \mathcal{F}_\sigma[P_\sigma(i)]$ can only decrease with time (Eq.~\ref{eq:generalised_entropy_change}), producing a non-equilibrium distribution  necessarily requires a higher initial generalised free energy than producing an equilibrium distribution. In other words, producing the non-equilibrium distribution has a greater minimal resource cost. 

Recent work has analysed how higher initial free energies allow self-assembly of non-equilibrium structures \cite{Sartori2015,Nguyen2016}. Intriguingly, these non-equilibrium assemblies can have a lower density of structural defects  than in equilibrium \cite{Sartori2015}, suggesting a possible alternative to optimising assembly by encouraging the approach to equilibrium. Indeed, it has recently been shown that a related process, the production of polymer copies that persist after separation from their templates (as occurs in replication, transcription and translation in cells)  is inherently an exercise in producing out-of-equilibrium structures. In this context, an equilibrium output is unrelated to its template, and so a high free energy initial state is required to give any accuracy at all
\cite{Ouldridge_copy_2016}.

The ability to interpret the properties of fluctuating trajectories on a thermodynamic level has also proven useful in understanding biochemical systems.
Molecular systems undergoing non-equilibrium stress-induced transitions in experiment can be understood using extensions to the above theory incorporating the application of external work \cite{Collin2005,Engel2014}. On a more theoretical level, an extremely common approach is to use imbalances in probability flows to infer the entropy generation (or free-energy consumption) of functional molecular networks \cite{Mehta2012,Lan2012,Barato2015,Pietzonka2016}. The entropy cost is paid by the consumption of a (typically implicit) molecular fuel molecule. Thus the resource cost of various molecular motifs, performing sensing, adaptive information-processing, timekeeping and force-generations, can be estimated and any trade-offs between performance and cost explored.  {An additional tool that has emerged within this field is the ``thermodynamic uncertainty relation" \cite{Barato2015,Pietzonka2016}, which imposes free-energy consumption bounds on the variability of  processes that operate cyclically, as many molecular systems do. Fundamentally, this uncertainty relation is based on physical constraints imposed on system dynamics that are not captured by the fluctuation relation (Eq.~\ref{eq:FR})  \cite{Pietzonka2016b}.}

One important feature of stochastic thermodynamics is the ability to relate the relative probabilities of entire trajectories and their microscopic reverses to entropy generation (Eq.~\ref{eq:db2}).  This allows the simultaneous analysis and comparison of the entropy cost of  distinct trajectories that move between macrostates via different pathways of different lengths. England has used this formalism to argue for a minimal bound on the entropy generation of replicators related to the overall birth and death rates \cite{England2013}. Similarly, by considering the entropy generated by simple and more complex activation pathways, it was recently demonstrated that single-step activation is optimally efficient for signal-propagating push-pull networks of the type shown in Fig.~\ref{fig:pushpull}\,(b) \cite{Ouldridge_copy_2015}.\\

\section{Conclusions}
\label{sec:conc}
In this pedagogical perspective, we have attempted to outline the manner in which molecular thermodynamics impacts the behaviour of both natural and artificial biochemical systems. Starting from the basic equilibrium statistical mechanics of dilute molecular systems, we introduced the concept of biochemical macrostates and the free energy of a macrostate. We then discussed the stochastic kinetics of molecular systems, and the constraints on kinetics implied by the macrostate free energies and detailed balance - which are strong, but insufficient to fully specify dynamics. Finally, we discussed how stochastic thermodynamics allows for a consistent thermodynamic interpretation of the stochastic evolution of a far-from-equilibrium system, both at the level of individual trajectories and the probability distribution as a whole.

Simultaneously, we have tried to highlight the relevance of these concepts to molecular systems that are typically studied and engineered. Thermodynamic ideas dictate both fundamental principles and general limits of what is possible, as well as contributing to detailed mechanistic understaning and design. In particular, defining biochemical macrostates allows for modelling of biochemical equilibria, and the state of a self-assembling system that would be expected in the limit of infinitely long time. In finite time, however, a system may not even come close to equilibrium -- it is therefore important to understand kinetics.  Kinetics are also relevant in systems that do not reach equilibrium because they are coupled  to a supply of molecular fuel. The operation of such systems, which  are widespread in nature but currently less common in artificial contexts, can be understood from the overriding tendency of the fuel molecules to relax towards their equilibrium. Finally, by defining entropies and free energies of trajectories and non-equilibrium distributions, stochastic thermodynamics allows for an analysis of the resource cost of various far-from-equilibrium processes that are important to both natural and artificial systems. 

At the same time, we have emphasised underlying assumptions that are often given insufficient consideration, and common misconceptions. For example, species concentration is an important component in the stability of self-assembling structures; a ``melting temperature" should never be defined without reference to concentration, and standard free energies are not directly meaningful at reasonable strand concentrations.  Markov models at the level of biochemical macrostates are only appropriate if the macrostates are well-defined so that transitions between them are rare events. Kinetic models of systems that are not subject to driving should exhibit detailed balance in equilibrium. This fact is sometimes ignored without careful consideration of the consequences; the assumption of perfect irreversibility can have major effects, depending on the context. Finally, catalysts can only influence the yield of downstream substrates if catalysis is coupled to a supply of non-equilibrium fuel.

The flow of understanding isn't solely in one direction, however. Biomolecular systems can also contribute to the understanding of fundamental thermodynamics. Most obviously, although stochastic thermodynamics wasn't developed solely for molecular systems, it finds perhaps it most natural application there. Important processes  in both natural and artificial molecular systems involve single molecules undergoing large fluctuations that are essential to function. Living systems are, almost by definition, kept in a far from equilibrium state. Pursuing important questions in such systems will thus drive understanding of the underlying thermodynamic principles, and the techniques used to study them. For example, many authors are currently probing the deep connections between information theory and (stochastic) thermodynamics in a quest to understand the limits and capabilities of cellular sensing, signalling and adaption networks  \cite{govern2014,Barato2014,Ouldridge_copy_2015,Ito2014}.

Much of the difficulty in understanding fundamental thermodynamics stems from interpreting the results, rather than from mathematical complexity. This is partly because many of the concepts, beginning with entropy itself, are inherently abstract. As a result, debate persists about the proper interpretation of relatively fundamental  systems and concepts \cite{Maroney2005,Ladyman2007,Norton2011,Dunkel2014} -- debates that we have ignored in this perspective. Biomolecular systems, however, are inherently concrete -- even when modelled in a simplistic manner. They can thus serve to demystify the debate. As an example, there has been much recent interest in the possibility of designing  an engine that exploits 1s and 0s on a tape to do useful work (such as lifting a weight) \cite{Mandal2012,Mandal2013}. When the state of the tape is written as abstract 1s and 0s, the operation of such a device seems  almost magical. However, for the device to run, the 1s and 0s must interact with a motor in a specific way; they must be physical. And a physical instantiation of the system can be constructed from biomolecules -- at which point it becomes clear that the 1s and 0s correspond to two states of a molecular fuel molecule, and the motor is driven by an imbalance of fuel input {\it just like any other motor} \cite{McGrath2017}. The operation of the device is consistent with well-established laws of thermodynamics, rather than being particularly remarkable.

 By constraining ourselves to physically plausible models of molecular systems, it is also much clearer what is possible. It is consequently harder to accidentally invent an unaccounted-for ``Maxwell Demon" that performs an impossible task, or miss the requirements of a key step. Moreover, there is at least some sense in which the practical constraints of thought experiments become evident. A thermodynamic system can be designed on paper with arbitrary coupling between degrees of freedom, but when trying to instantiate it as a molecular system, even theoretically, the inherent trade-offs required  become more obvious \cite{McGrath2017}.

In particular, the majority of thermodynamic studies focus on systems in which changes are driven by intervention from an external experimenter, who does work on the system (this idea is briefly introduced in  Section~\ref{sec:fundamentals} and \ref{sec:thermo_reversibility}). From the perspective of stochastic thermodynamics, doing work corresponds to adjusting the system to change the energy of the microstates in a time-dependent manner. Whilst analysis in terms of external work is not wrong, it is again hard to understand what is fundamentally possible at zero cost, particularly for small systems. Generally, the actual physical mechanism by which work is applied is not considered; it is therefore unclear which work protocols are feasible, and whether work can actually be applied, recovered and stored efficiently \cite{Gopalkrishnan2015}, as is assumed. For example, recent experiments on the manipulation of single colloids actually implement ``work" protocols through a highly dissipative mechanism, the cost of which far exceeds any entropy generated by the motion of the colloid itself (which is typically the topic of interest) \cite{Berut2012,Jun2014}.  Similarly, in questions related to the minimal cost of operations on small systems, such as computing with a single bit, the cost of an experimenter deciding to implement each stage of the protocol is not obviously accounted for \cite{Gopalkrishnan2015}. 

By contrast, molecular systems are  usually thought of as running autonomously, without external work (although this is not always the case \cite{Ouldridge_copy_2015,Ouldridge_copy_2016}). Due to the ability of molecules to diffuse and interact selectively with multiple partners, complex behaviour can arise simply from a system initiated in some non-equilibrium state and left to evolve (perhaps with a constant supply of non-equilibrium fuel). In these autonomous systems, all costs are explicit, and all behaviours based on plausible chemical reactions. A major challenge for the future of thermodynamics lies in understanding the fundamental differences between the capabilities of such autonomous systems, and those systems in which outside manipulation is permitted. In turn, understanding these autonomous systems will potentially allow us to build microscopic devices of extremely high efficiency that genuinely approach the fundamental lower bounds on cost \cite{Ouldridge_copy_2015}.

\begin{acknowledgements}
The author wishes to acknowledge fruitful conversations with  P. R. ten Wolde, A. A. Louis, J. P. K. Doye, M. Gopalkrishnan and A. J. Turberfield.
\end{acknowledgements}


%
%

\end{document}